\documentclass[12pt,a4paper]{article}
\makeatletter
\def\eqnarray{%
\stepcounter{equation}%
\let\@currentlabel=\theequation
\global\@eqnswtrue
\global\@eqcnt\z@
\tabskip\@centering
\let\\=\@eqncr
$$\halign to \displaywidth\bgroup\@eqnsel\hskip\@centering
$\displaystyle\tabskip\z@{##}$&\global\@eqcnt\@ne
\hfil$\displaystyle{{}##{}}$\hfil
&\global\@eqcnt\tw@$\displaystyle\tabskip\z@{##}$\hfil
\tabskip\@centering&\llap{##}\tabskip\z@\cr}
\makeatother
\newcommand{\ket}[1]{{\vert{#1}\rangle}}

\newcommand{\fukuso}{{\mathbf C}}
\newcommand{\futon}{{\bf N}}

\begin{document}

\title{\sl Quantum Optical Construction of Generalized Pauli and 
Walsh--Hadamard Matrices in Three Level Systems}
\author{
  Kazuyuki FUJII 
  \thanks{E-mail address : fujii@yokohama-cu.ac.jp }\\
  Department of Mathematical Sciences\\
  Yokohama City University\\
  Yokohama, 236--0027\\
  Japan
  }
\date{}
\maketitle
%
%
%
%
\begin{abstract}
  A set of generators of generalized Pauli matrices play a crucial role in 
  quantum computation based on n level systems of an atom. 
  In this paper we show how to construct them by making use of Rabi 
  oscillations. We also construct the generalized Walsh--Hadamard matrix 
  in the case of three level systems and present some related problems.
\end{abstract}

\newpage

%
%
%
%

\section{Introduction}

Quantum Computation (or Computer) is a challenging task in this century 
for not only physicists but also mathematicians. 
See for example \cite{LPS} as a general introduction to it. 

Quantum Computation is in a usual understanding based on qubits which are 
based on two level systems (two energy levels or fundamental spins) of atoms, 
See \cite{AE}, \cite{MSIII}, \cite{BR}, \cite{C-HDG} as for general theory of 
two level systems. 

In a realistic image of Quantum Computer we need at least one hundred atoms. 
However then we meet a very severe problem called Decoherence which will 
destroy a superposition of quantum states in the process of unitary 
evolution of our system. At the present it is not easy to control Decoherence.
See for example \cite{WHZ} or recent \cite{MFr3} as an introduction.  

By the way, an atom has in general infinitely many energy levels, while 
in a qubit method we use only two energy ones. We should use this possibility 
to reduce a number of atoms. 
We use $n$ energy levels from the ground state (it is not realistic to take 
all energy levels into consideration at the same time). We call this 
$n$ level systems a qudit theory, see for example \cite{KF3}, \cite{KF5}, 
\cite{FHKW}, \cite{KuF}. 

In quantum computation based on a qudit theory the generalized Pauli matrices 
$\{\Sigma_{1},\ \Sigma_{3}\}$ and the generalized Walsh--Hadamard matrix $W$ 
play a central role, see \cite{CBKG}, \cite{BGS}, \cite{KBB}, \cite{KF4}. 
Therefore we must first of all construct them. 

In a qubit case we need Rabi oscillations to construct quantum logic gates. 
See \cite{KF1} as a simple introduction to quantum logic gates. 
Similarly we also need Rabi oscillations to construct them in a qudit space. 
However a general theory of Rabi oscillations in $n$ level systems has not 
been developed enough as far as we know. For three level systems see 
\cite{MSIII}, \cite{BR}, \cite{MFr5} and \cite{FHKW}.
 
Therefore we develop such a theory in this paper and construct the 
generalized Pauli matrices and Walsh--Hadamard matrix by making use of 
Rabi oscillations in three level systems. 

In this paper we assume the rotating wave approximation (RWA) in our model 
from the beginning, otherwise we cannot solve the model. However there is 
no problem on the approximation in the weak coupling regime. 
We note that some problems will appear when using this approximation 
in the strong coupling regime.

\section{Two Level System}

In this section we make a review of Rabi oscillations (or coherent 
oscillations) in two level systems of an atom and apply them to constructing 
quantum logic gates in quantum computation. 
For a quantum version of the Rabi oscillations see for example \cite{C-HDG}, 
\cite{MFr2}, \cite{KF2}.

\subsection{General Theory}

Let $\{\sigma_{1}, \sigma_{2}, \sigma_{3}\}$ be Pauli matrices : 
\begin{equation}
\label{eq:pauli}
\sigma_{1} = 
\left(
  \begin{array}{cc}
    0& 1 \\
    1& 0
  \end{array}
\right), \quad 
\sigma_{2} = 
\left(
  \begin{array}{cc}
    0& -i \\
    i& 0
  \end{array}
\right), \quad 
\sigma_{3} = 
\left(
  \begin{array}{cc}
    1& 0 \\
    0& -1
  \end{array}
\right), 
\end{equation}
and we set 
\[
\sigma_{+}=\frac{1}{2}(\sigma_{1}+i\sigma_{2})=
\left(
  \begin{array}{cc}
    0& 1 \\
    0& 0
  \end{array}
\right), \quad 
\sigma_{-}=\frac{1}{2}(\sigma_{1}-i\sigma_{2})=
\left(
  \begin{array}{cc}
    0& 0 \\
    1& 0
  \end{array}
\right). 
\]

Let us consider an atom with $2$ energy levels $E_{0}$ and $E_{1}$ (
$E_{1} > E_{0}$). 
Its Hamiltonian is in the diagonal form given as 
\begin{equation}
H_{0}=
\left(
  \begin{array}{cc}
    E_{0}& 0 \\
    0& E_{1}
  \end{array}
\right).
\end{equation}
This is rewritten as 
\[
H_{0}=
E_{0}
\left(
  \begin{array}{cc}
    1& 0 \\
    0& 1
  \end{array}
\right)+
\left(
  \begin{array}{cc}
    0& 0           \\
    0& E_{1}-E_{0}
  \end{array}
\right)
=
E_{0}{\bf 1}_{2}+
\left(
  \begin{array}{cc}
    0& 0       \\
    0& \Delta
  \end{array}
\right)
\]
where $\Delta=E_{1}-E_{0}$ is an energy difference. 

We consider an atom with two energy levels which interacts with external 
(periodic) field with $g\mbox{cos}(\omega t+\phi)$. 
In the following we set $\hbar=1$ for simplicity. 
The Hamiltonian in the dipole approximation is given by 
\begin{equation}
\label{eq:2-full-hamiltonian}
H=H_{0}+g \mbox{cos}(\omega t+\phi)\sigma_{1}, 
\end{equation}
where $\omega$ is the frequency of the external field, $g$ the coupling 
constant between the external field and the atom. This model is complicated 
enough, see \cite{AE}, \cite{MFr1}. 

In the following we assume the rotating wave approximation (which neglects 
the fast oscillating terms), namely 
\[
\mbox{cos}(\omega t+\phi)
=\frac{1}{2}(\mbox{e}^{i(\omega t+\phi)}+\mbox{e}^{-i(\omega t+\phi)})
=\frac{1}{2}\mbox{e}^{i(\omega t+\phi)}(1+\mbox{e}^{-2i(\omega t+\phi)})
\approx \frac{1}{2}\mbox{e}^{i(\omega t+\phi)},
\]
and 
\[
\mbox{cos}(\omega t+\phi)\sigma_{1}
=
\left(
  \begin{array}{cc}
                     0 & \mbox{cos}(\omega t+\phi) \\
    \mbox{cos}(\omega t+\phi)& 0
  \end{array}
\right)
\approx 
\frac{1}{2}
\left(
  \begin{array}{cc}
                       0 & \mbox{e}^{i(\omega t+\phi)} \\
    \mbox{e}^{-i(\omega t+\phi)}& 0
  \end{array}
\right), 
\]
therefore the Hamiltonian is given by 
\begin{eqnarray}
\label{eq:sub-hamiltonian}
H&=&E_{0}{\bf 1}_{2}+
\frac{\Delta}{2}\left({\bf 1}_{2}-\sigma_{3}\right)+\frac{g}{2}
\left(\mbox{e}^{i(\omega t+\phi)}\sigma_{+}+
\mbox{e}^{-i(\omega t+\phi)}\sigma_{-}\right)   \nonumber \\
&\equiv& 
E_{0}{\bf 1}_{2}+
\frac{\Delta}{2}\left({\bf 1}_{2}-\sigma_{3}\right)+
g\left(\mbox{e}^{i(\omega t+\phi)}\sigma_{+}+
\mbox{e}^{-i(\omega t+\phi)}\sigma_{-}\right)   \nonumber 
\end{eqnarray}
by the redefinition of $g$ ($g/2 \longrightarrow g$). 
It is explicitly 
\begin{equation}
\label{eq:reduction-matrix}
H=E_{0}{\bf 1}_{2}+
\left(
  \begin{array}{cc}
    0& g\mbox{e}^{i(\omega t+\phi)}         \\
    g\mbox{e}^{-i(\omega t+\phi)}& \Delta
  \end{array}
\right). 
\end{equation}

We would like to solve the Schr{\" o}dinger equation 
\begin{equation}
\label{eq:schrodinger-equation}
i\frac{d}{dt}\Psi=H\Psi.
\end{equation}
For that let us decompose $H$ in (\ref{eq:reduction-matrix}) into 
\begin{equation}
\label{eq:decomposition}
\left(
  \begin{array}{cc}
    0& g\mbox{e}^{i(\omega t+\phi)}         \\
    g\mbox{e}^{-i(\omega t+\phi)}& \Delta
  \end{array}
\right)
=
\left(
  \begin{array}{cc}
    1&         \\
     & \mbox{e}^{-i(\omega t+\phi)}
  \end{array}
\right)
\left(
  \begin{array}{cc}
    0& g         \\
    g& \Delta
  \end{array}
\right)
\left(
  \begin{array}{cc}
    1&         \\
     & \mbox{e}^{i(\omega t+\phi)}
  \end{array}
\right),
\end{equation}
so if we set 
\begin{equation}
\label{eq:transformation}
\Phi=\mbox{e}^{itE_{0}}
\left(
  \begin{array}{cc}
    1&         \\
     & \mbox{e}^{i(\omega t+\phi)}
  \end{array}
\right)
\Psi
\quad \Longleftrightarrow \quad 
\Psi=\mbox{e}^{-itE_{0}}
\left(
  \begin{array}{cc}
    1&         \\
     & \mbox{e}^{-i(\omega t+\phi)}
  \end{array}
\right)
\Phi
\end{equation}
then it is not difficult to see 
\begin{equation}
\label{eq:schrodinger-equation-reduction}
i\frac{d}{dt}\Phi=
\left(
  \begin{array}{cc}
    0& g              \\
    g& \Delta-\omega 
  \end{array}
\right)
\Phi,
\end{equation}
which is easily solved. For simplicity we set the resonance condition 
\begin{equation}
\label{eq:resonance}
  \Delta=\omega,
\end{equation}
then the solution of (\ref{eq:schrodinger-equation-reduction}) is 
\[
\Phi(t)=
\mbox{exp}
\left\{-igt
\left(
  \begin{array}{cc}
    0& 1   \\
    1& 0 
  \end{array}
\right)
\right\}
\Phi(0)=
\left(
  \begin{array}{cc}
    \mbox{cos}(gt)& -i\mbox{sin}(gt)   \\
    -i\mbox{sin}(gt)& \mbox{cos}(gt) 
  \end{array}
\right)
\Phi(0).
\]
As a result, the solution of the equation (\ref{eq:schrodinger-equation}) is 
given as 
\begin{eqnarray}
\Psi(t)
&=&\mbox{e}^{-itE_{0}}
\left(
  \begin{array}{cc}
    1&         \\
     & \mbox{e}^{-i(\omega t+\phi)}
  \end{array}
\right)
\Phi(t)     \nonumber \\
&=&\mbox{e}^{-itE_{0}}
\left(
  \begin{array}{cc}
    1&         \\
     & \mbox{e}^{-i(\omega t+\phi)}
  \end{array}
\right)
\left(
 \begin{array}{cc}
   \mbox{cos}(gt)  & -i\mbox{sin}(gt)   \\
   -i\mbox{sin}(gt)& \mbox{cos}(gt) 
 \end{array}
\right)
\Phi(0)
\end{eqnarray}
by (\ref{eq:transformation}). If we choose 
$
\Phi(0)=\left(1,\ 0\right)^{T}
$
 as an initial condition, then 
\begin{equation}
\Psi(t)=\mbox{e}^{-itE_{0}}
\left(
  \begin{array}{c}
    \mbox{cos}(gt)                             \\
  -i\mbox{e}^{-i(\omega t+\phi)}\mbox{sin}(gt)
  \end{array}
\right).
\end{equation}
This is a well--known model of the Rabi oscillation (or coherent oscillation). 

For the latter use we set 
\begin{equation}
U(t,0)=
\mbox{e}^{-itE_{0}}
\left(
  \begin{array}{cc}
    1&         \\
     & \mbox{e}^{-i(\omega t+\phi)}
  \end{array}
\right)
\left(
 \begin{array}{cc}
   \mbox{cos}(gt)  & -i\mbox{sin}(gt)   \\
   -i\mbox{sin}(gt)& \mbox{cos}(gt) 
 \end{array}
\right)
\end{equation}
and 
\begin{equation}
\label{eq:unitary-u}
U(t_{f},t_{i})=U(t_{f}-t_{i},0)
\end{equation}
because $\Delta=\omega$ due to the resonance condition (\ref{eq:resonance}). 
Moreover, for $g=0$ (no interaction with external field), then we have 
\begin{equation}
\label{eq:unitary-v}
V(t_{f},t_{i})=
\mbox{e}^{-i(t_{f}-t_{i})E_{0}}
\left(
 \begin{array}{cc}
   1&                       \\
    & \mbox{e}^{-i\Delta(t_{f}-t_{i})}
 \end{array}
\right).
\end{equation}

We construct several quantum logic gates useful by combining $U(t_{f},t_{i})$ 
and $V(t_{f},t_{i})$ for appropriate $t_{i} < t_{f}$ in the following 
sections. Then we will change a time $t$ and a phase $\phi$ as free 
parameters (we don't change the coupling constant $g$).

\subsection{Quantum Logic Gates}

As an exercise we would like to construct useful unitary matrices 
\begin{equation}
\sigma_{1} = 
\left(
  \begin{array}{cc}
    0& 1 \\
    1& 0
  \end{array}
\right), \quad 
\sigma_{\theta} = 
\left(
  \begin{array}{cc}
    1& 0                  \\
    0& \mbox{e}^{i\theta}
  \end{array}
\right), \quad 
W = \frac{1}{\sqrt{2}}
\left(
  \begin{array}{cc}
    1&  1 \\
    1& -1
  \end{array}
\right) 
\end{equation}
for any $\theta$ explicitly. 

First we choose $t_{1}$ as $t_{1}=\frac{3\pi}{2\Delta}$ satisfying 
\[
V(t_{1},0)
=
\mbox{e}^{-iE_{0}t_{1}}
\left(
 \begin{array}{cc}
   1&                           \\
    & \mbox{e}^{-i\Delta t_{1}}
 \end{array}
\right)
=
\mbox{e}^{-iE_{0}t_{1}}
\left(
  \begin{array}{cc}
    1& 0 \\
    0& i
  \end{array}
\right), 
\]
and next choose $t_{2}-t_{1}$ as $t_{2}-t_{1}=\frac{\pi}{2g}$ satisfying 
\begin{eqnarray}
U(t_{2},t_{1})&=&
\mbox{e}^{-iE_{0}(t_{2}-t_{1})}
\left(
  \begin{array}{cc}
    1&         \\
     & \mbox{e}^{-i\{\Delta(t_{2}-t_{1})+\phi\}}
  \end{array}
\right)
\left(
 \begin{array}{cc}
   \mbox{cos}(g(t_{2}-t_{1}))& -i\mbox{sin}(g(t_{2}-t_{1}))   \\
   -i\mbox{sin}(g(t_{2}-t_{1}))& \mbox{cos}(g(t_{2}-t_{1})) 
 \end{array}
\right)            \nonumber \\
&=&
\mbox{e}^{-iE_{0}(t_{2}-t_{1})}
\left(
  \begin{array}{cc}
    1&         \\
     & \mbox{e}^{-i\{\Delta(t_{2}-t_{1})+\phi\}}
  \end{array}
\right)
\left(
 \begin{array}{cc}
     & -i   \\
   -i& 
 \end{array}
\right).         \nonumber 
\end{eqnarray}
Then 
\[
U(t_{2},t_{1})V(t_{1},0)
=
\mbox{e}^{-iE_{0}t_{2}}
\left(
  \begin{array}{cc}
    1&         \\
     & \mbox{e}^{-i\{\Delta(t_{2}-t_{1})+\phi\}}
  \end{array}
\right)
\left(
 \begin{array}{cc}
     & 1   \\
   -i& 
 \end{array}
\right).   
\]
Moreover multiplying $V(t_{3},t_{2})$ from the left 
\[
V(t_{3},t_{2})U(t_{2},t_{1})V(t_{1},0)
=
\mbox{e}^{-iE_{0}t_{3}}
\left(
  \begin{array}{cc}
    1&         \\
     & \mbox{e}^{-i\{\Delta(t_{3}-t_{1})+\phi\}}
  \end{array}
\right)
\left(
 \begin{array}{cc}
     & 1   \\
   -i& 
 \end{array}
\right)   
\]
Here we choose $t_{3}=2k\pi/E_{0}\ ( > t_{2} > t_{1})$ as 
$\mbox{e}^{-iE_{0}t_{3}}=1$ and the phase $\phi$ as 
$\mbox{e}^{-i\{\Delta(t_{3}-t_{1})+\phi\}}=i$, so that 
\[
V(t_{3},t_{2})U(t_{2},t_{1})V(t_{1},0)
=
\left(
  \begin{array}{cc}
    1&     \\
     & i
  \end{array}
\right)
\left(
 \begin{array}{cc}
     & 1   \\
   -i& 
 \end{array}
\right) 
=
\left(
 \begin{array}{cc}
     & 1   \\
    1& 
 \end{array}
\right)=\sigma_{1}.   
\]

\vspace{5mm}
Next we construct $\sigma_{\theta}$. We choose $t_{1}$ as $t_{1}=2\pi/g$ 
\[
U(t_{1},0)=\mbox{e}^{-iE_{0}t_{1}}
\left(
  \begin{array}{cc}
    1&         \\
     & \mbox{e}^{-i(\Delta t_{1}+\phi)}
  \end{array}
\right)
\left(
 \begin{array}{cc}
   \mbox{cos}(gt_{1})& -i\mbox{sin}(gt_{1})   \\
   -i\mbox{sin}(gt_{1})& \mbox{cos}(gt_{1}) 
 \end{array}
\right)
=
\mbox{e}^{-iE_{0}t_{1}}
\left(
  \begin{array}{cc}
    1&         \\
     & \mbox{e}^{-i(\Delta t_{1}+\phi)}
  \end{array}
\right).
\]
Then multiplying $V(t_{2},t_{1})$ from the left 
\[
V(t_{2},t_{1})U(t_{1},0)
=
\mbox{e}^{-iE_{0}t_{2}}
\left(
  \begin{array}{cc}
    1&                                   \\
     & \mbox{e}^{-i(\Delta t_{2}+\phi)}
  \end{array}
\right).
\]
Here we choose $t_{2}=2k\pi/E_{0}\ (> t_{1})$ for some $k \in \futon$ and 
the phase $\phi$ as 
$\mbox{e}^{-i(\Delta t_{2}+\phi)}=\mbox{e}^{i\theta}$ for any $\theta$, 
so that 
\[
V(t_{2},t_{1})U(t_{1},0)
=
\left(
  \begin{array}{cc}
    1&         \\
     & \mbox{e}^{i\theta}
  \end{array}
\right)=\sigma_{\theta}.
\]
In particular we obtain 
\[
\left(
  \begin{array}{cc}
    1&   \\
     & i
  \end{array}
\right), \quad 
\sigma_{3}=
\left(
  \begin{array}{cc}
    1&    \\
     & -1
  \end{array}
\right), \quad 
\left(
  \begin{array}{cc}
    1&   \\
     & -i
  \end{array}
\right)
\]
for $\theta=\pi/2,\ \pi,\ 3\pi/2$ respectively.

\vspace{5mm}
Lastly we construct the Walsh--Hadamard matrix which plays a central role 
in quantum computation based on qubits. For $V(t_{1},0)$, 
$V(t_{3},t_{2})$ with $t_{1}=3\pi/2\Delta$ and $t_{3}-t_{2}=3\pi/2\Delta$ 
\[
V(t_{1},0)
=
\mbox{e}^{-iE_{0}t_{1}}
\left(
  \begin{array}{cc}
    1& 0 \\
    0& i
  \end{array}
\right), \quad 
V(t_{3},t_{2})
=
\mbox{e}^{-iE_{0}(t_{3}-t_{2})}
\left(
  \begin{array}{cc}
    1& 0 \\
    0& i
  \end{array}
\right)
\]
and $U(t_{2},t_{1})$ with $t_{2}-t_{1}=\pi/4g$
\[
U(t_{2},t_{1})
=
\mbox{e}^{-iE_{0}(t_{2}-t_{1})}
\left(
  \begin{array}{cc}
    1&         \\
     & \mbox{e}^{-i\{\Delta(t_{2}-t_{1})+\phi\}}
  \end{array}
\right)
\left(
  \begin{array}{cc}
    \frac{1}{\sqrt{2}} & \frac{-i}{\sqrt{2}}  \\
    \frac{-i}{\sqrt{2}}& \frac{1}{\sqrt{2}} 
  \end{array}
\right), 
\]
we have 
\[
V(t_{3},t_{2})U(t_{2},t_{1})V(t_{1},0)
=
\mbox{e}^{-iE_{0}t_{3}}
\left(
  \begin{array}{cc}
    1&         \\
     & \mbox{e}^{-i\{\Delta(t_{2}-t_{1})+\phi\}}
  \end{array}
\right)
\left(
  \begin{array}{cc}
    \frac{1}{\sqrt{2}}& \frac{1}{\sqrt{2}}  \\
    \frac{1}{\sqrt{2}}& -\frac{1}{\sqrt{2}} 
  \end{array}
\right).
\]
We cannot remove the phase $\mbox{e}^{-iE_{0}t_{3}}$, so we multiply the above 
by $V(t_{4},t_{3})$ to obtain 
\[
V(t_{4},t_{3})V(t_{3},t_{2})U(t_{2},t_{1})V(t_{1},0)
=
\mbox{e}^{-iE_{0}t_{4}}
\left(
  \begin{array}{cc}
    1&         \\
     & \mbox{e}^{-i\{\Delta(t_{4}-t_{3}+t_{2}-t_{1})+\phi\}}
  \end{array}
\right)
\left(
  \begin{array}{cc}
    \frac{1}{\sqrt{2}}& \frac{1}{\sqrt{2}}  \\
    \frac{1}{\sqrt{2}}& -\frac{1}{\sqrt{2}} 
  \end{array}
\right).
\]
By choosing $t_{4}=2k\pi/E_{0}$ for some $k \in \futon$ and the phase $\phi$ 
as $\mbox{e}^{-i\{\Delta(t_{4}-t_{3}+t_{2}-t_{1})+\phi\}}=1$ 
we finally obtain 
\[
V(t_{4},t_{3})V(t_{3},t_{2})U(t_{2},t_{1})V(t_{1},0)
=
\frac{1}{\sqrt{2}}
\left(
  \begin{array}{cc}
    1& 1  \\
    1& -1
  \end{array}
\right)=W. 
\]

\section{Three Level System}

In this section we consider an atom with three energy levels 
$\{\ket{0},E_{0}\},\ \{\ket{1},E_{1}\},\ \{\ket{2},E_{2}\}$ which interacts 
with external fields. 
As for the external fields we use laser fields with frequencies equal to 
energy differences of the atom. Now we set 
\[
\Delta_{1}=E_{1}-E_{0}, \quad \Delta_{2}=E_{2}-E_{0} 
\Longleftrightarrow 
E_{1}-E_{0}=\Delta_{1},\quad E_{2}-E_{1}=\Delta_{2}-\Delta_{1}
\]
for the latter convenience and assume $E_{1}-E_{0} > E_{2}-E_{1}$. 
See the following picture : 

\vspace{5mm}
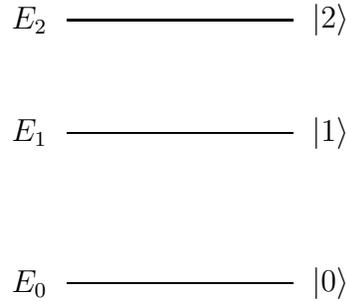
\begin{center}
\setlength{\unitlength}{1mm}   %
\begin{picture}(60,60)(0,0)
\put(5,50){\makebox(10,10)[c]{$E_2$}}
\put(15,55){\line(1,0){30}}
\put(45,50){\makebox(10,10)[c]{$|2\rangle$}}
\put(5,35){\makebox(10,10)[c]{$E_1$}}
\put(15,40){\line(1,0){30}}
\put(45,35){\makebox(10,10)[c]{$|1\rangle$}}
\put(5,15){\makebox(10,10)[c]{$E_0$}}
\put(15,20){\line(1,0){30}}
\put(45,15){\makebox(10,10)[c]{$|0\rangle$}}
\put(30,8){\circle*{3}}
\end{picture}
\end{center}
\vspace{-5mm}

Under the following conditions depending on external fields we solve 
the Schr{\" o}dinger equations and obtain unitary transformations. 
These transformations will play an central role in constructing the 
generalized Pauli matrices and generalized Walsh--Hadamard matrix 
in the next section.

\subsection{Unitary Transformation of type 0}

First we consider an atom with no interaction with external fields. 
The Hamiltonian that we are treating is 
\begin{equation}
\label{eq:hamiltonian-0}
H_{0}=
\left(
\begin{array}{ccc}
   E_{0} &       &        \\
         & E_{1} &        \\
         &       & E_{2}
  \end{array}
\right)
=E_{0}{\bf 1}_{3}+
\left(
  \begin{array}{ccc}
   0  &             &             \\
      & \Delta_{1}  &             \\
      &             & \Delta_{2}
  \end{array}
\right).
\end{equation}

We would like to solve the Schr{\" o}dinger equation 
\begin{equation}
\label{eq:schrodinger-equation-0}
i\frac{d}{dt}\Psi=H_{0}\Psi.
\end{equation}
The solution is easily obtained to be 
\begin{equation}
\label{eq:full-solution-0}
\Psi(t)=\mbox{e}^{-itE_{0}}
\left(
  \begin{array}{ccc}
  1 &                          &                            \\
    & \mbox{e}^{-it\Delta_{1}} &                            \\
    &                          & \mbox{e}^{-it\Delta_{2}}
  \end{array}
\right)\Psi(0).
\end{equation}

For the latter use we set 
\begin{equation}
\label{eq:unitary-0}
U_{0}(t,0)=\mbox{e}^{-itE_{0}}
\left(
  \begin{array}{ccc}
  1 &                          &                            \\
    & \mbox{e}^{-it\Delta_{1}} &                            \\
    &                          & \mbox{e}^{-it\Delta_{2}}
  \end{array}
\right).
\end{equation}

\subsection{Unitary Transformation of type I}

We consider the following case. That is, we shoot the atom with a laser field 
having the frequency equal to the energy difference $E_{1}-E_{0}$. 
See the following picture.

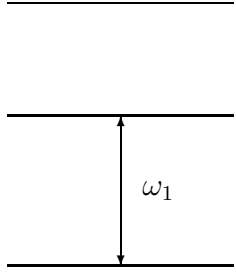
\vspace{5mm}
\begin{center}
\setlength{\unitlength}{1mm}   %
%
\end{center}
\vspace{-15mm}

The Hamiltonian that we are treating is 
\begin{equation}
\label{eq:hamiltonian-1}
H_{I}=
\left(
  \begin{array}{ccc}
   E_{0} & g\mbox{e}^{i(\phi_{1}+\omega_{1}t)} &   \\
   g\mbox{e}^{-i(\phi_{1}+\omega_{1}t)} & E_{1} &  \\
     &  & E_{2}
  \end{array}
\right)
=E_{0}{\bf 1}_{3}+
\left(
  \begin{array}{ccc}
   0  & g\mbox{e}^{i(\phi_{1}+\omega_{1}t)} &      \\
   g\mbox{e}^{-i(\phi_{1}+\omega_{1}t)} & \Delta_{1} &  \\
     &  & \Delta_{2}
  \end{array}
\right).
\end{equation}

We would like to solve the Schr{\" o}dinger equation 
\begin{equation}
\label{eq:schrodinger-equation-1}
i\frac{d}{dt}\Psi=H_{I}\Psi.
\end{equation}

If we note the following decomposition 
\begin{eqnarray}
&&\left(
  \begin{array}{ccc}
   0  & g\mbox{e}^{i(\phi_{1}+\omega_{1}t)} &            \\
   g\mbox{e}^{-i(\phi_{1}+\omega_{1}t)} & \Delta_{1} &   \\
     &  & \Delta_{2}
  \end{array}
\right)
=               \nonumber \\
&&\left(
  \begin{array}{ccc}
  1 & &                                          \\
    & \mbox{e}^{-i(\phi_{1}+\omega_{1}t)} &      \\
    & & \mbox{e}^{-it\Delta_{2}}
  \end{array}
\right)
\left(
  \begin{array}{ccc}
    0 & g &     \\
    g & \Delta_{1} &    \\
      &   & \Delta_{2}
  \end{array}
\right)
\left(
  \begin{array}{ccc}
  1 & &                                         \\
    & \mbox{e}^{i(\phi_{1}+\omega_{1}t)} &      \\
    & & \mbox{e}^{it\Delta_{2}}
  \end{array}
\right)   
\end{eqnarray}
and define 
\begin{equation}
\Phi=\mbox{e}^{itE_{0}}
\left(
  \begin{array}{ccc}
  1 & &                                           \\
    & \mbox{e}^{i(\phi_{1}+\omega_{1}t)} &        \\
    & & \mbox{e}^{it\Delta_{2}}
  \end{array}
\right)\Psi 
\end{equation}
, then it is not difficult to see 
\begin{equation}
i\frac{d}{dt}\Phi=
\left(
  \begin{array}{ccc}
   0 & g &                       \\
   g & \Delta_{1}-\omega_{1} &   \\
     &  & 0
  \end{array}
\right)\Phi
\equiv 
\tilde{H}_{I}\Phi.
\end{equation}
Here we set the resonance condition 
\begin{equation}
\label{eq:resonance-1}
  \Delta_{1}=\omega_{1},
\end{equation}
so the solution is easily obtained to be 
\begin{equation}
\Phi(t)=\mbox{exp}(-it\tilde{H}_{I})\Phi(0)=
\left(
  \begin{array}{ccc}
   \mbox{cos}(gt) & -i\mbox{sin}(gt) &   \\
   -i\mbox{sin}(gt) & \mbox{cos}(gt) &   \\
     &  &  1
  \end{array}
\right)\Phi(0).
\end{equation}
As a result the solution we are looking for is 
\begin{equation}
\label{eq:full-solution-1}
\Psi(t)=\mbox{e}^{-itE_{0}}
\left(
  \begin{array}{ccc}
  1 & &                                        \\
    & \mbox{e}^{-i(\phi_{1}+\omega_{1}t)} &    \\
    & & \mbox{e}^{-it\Delta_{2}}
  \end{array}
\right)
\left(
  \begin{array}{ccc}
   \mbox{cos}(gt) & -i\mbox{sin}(gt) &   \\
   -i\mbox{sin}(gt) & \mbox{cos}(gt) &   \\
     &  &  1
  \end{array}
\right)\Phi(0).
\end{equation}

For the latter use we set 
\begin{equation}
\label{eq:unitary-1}
U_{1}(t,0)=\mbox{e}^{-itE_{0}}
\left(
  \begin{array}{ccc}
  1 & &                                        \\
    & \mbox{e}^{-i(\phi_{1}+\omega_{1}t)} &    \\
    & & \mbox{e}^{-it\Delta_{2}}
  \end{array}
\right)
\left(
  \begin{array}{ccc}
   \mbox{cos}(gt) & -i\mbox{sin}(gt) &   \\
   -i\mbox{sin}(gt) & \mbox{cos}(gt) &   \\
     &  &  1
  \end{array}
\right).
\end{equation}

\subsection{Unitary Transformation of type II}

We consider the following case. That is, we shoot the atom with a laser field 
having the frequency equal to the energy difference $E_{2}-E_{1}$. 
See the following picture.

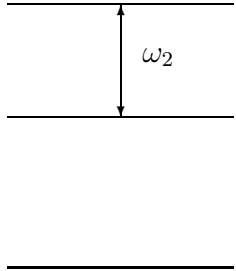
\vspace{5mm}
\begin{center}
\setlength{\unitlength}{1mm}   %
%
\end{center}
\vspace{-15mm}

The Hamiltonian that we are treating is 
\begin{equation}
\label{eq:hamiltonian-2}
H_{II}=
\left(
  \begin{array}{ccc}
   E_{0} &       &                                          \\
         & E_{1} & g\mbox{e}^{i(\phi_{2}+\omega_{2}t)}      \\
         & g\mbox{e}^{-i(\phi_{2}+\omega_{2}t)} & E_{2}
  \end{array}
\right)
=E_{1}{\bf 1}_{3}+
\left(
  \begin{array}{ccc}
   -\Delta_{1} &  &                                                  \\
               &  0 & g\mbox{e}^{i(\phi_{2}+\omega_{2}t)}            \\
     & g\mbox{e}^{-i(\phi_{2}+\omega_{2}t)} & \Delta_{2}-\Delta_{1}
  \end{array}
\right).
\end{equation}

We would like to solve the Schr{\" o}dinger equation 
\begin{equation}
\label{eq:schrodinger-equation-2}
i\frac{d}{dt}\Psi=H_{II}\Psi.
\end{equation}

If we note the following decomposition 
\begin{eqnarray}
&&\left(
  \begin{array}{ccc}
   -\Delta_{1} &  &                                                  \\
     & 0 & g\mbox{e}^{i(\phi_{2}+\omega_{2}t)}                       \\
     & g\mbox{e}^{-i(\phi_{2}+\omega_{2}t)}  & \Delta_{2}-\Delta_{1}
  \end{array}
\right)
=               \nonumber \\
&&\left(
  \begin{array}{ccc}
  \mbox{e}^{it\Delta_{1}} &   &                 \\
    & 1 &                                        \\
    &   & \mbox{e}^{-i(\phi_{2}+\omega_{2}t)} 
  \end{array}
\right)
\left(
  \begin{array}{ccc}
    -\Delta_{1} &  &                 \\
      & 0 & g                        \\
      & g & \Delta_{2}-\Delta_{1}
  \end{array}
\right)
\left(
  \begin{array}{ccc}
  \mbox{e}^{-it\Delta_{1}} &   &                \\
    & 1 &                                       \\
    &   & \mbox{e}^{i(\phi_{2}+\omega_{2}t)} 
  \end{array}
\right)   
\end{eqnarray}
and define 
\begin{equation}
\Phi=\mbox{e}^{itE_{1}}
\left(
  \begin{array}{ccc}
  \mbox{e}^{-it\Delta_{1}} &   &               \\
    & 1 &                                      \\
    &   & \mbox{e}^{i(\phi_{2}+\omega_{2}t)} 
  \end{array}
\right)\Psi 
\end{equation}
, then it is not difficult to see 
\begin{equation}
i\frac{d}{dt}\Phi=
\left(
  \begin{array}{ccc}
   0 &   &                                     \\
     & 0 & g                                   \\
     & g & \Delta_{2}-\Delta_{1}-\omega_{2} 
  \end{array}
\right)\Phi
\equiv 
\tilde{H}_{II}\Phi.
\end{equation}
Here we set the resonance condition 
\begin{equation}
\label{eq:resonance-2}
  \Delta_{2}-\Delta_{1}=\omega_{2},
\end{equation}
so the solution is easily obtained to be 
\begin{equation}
\Phi(t)=\mbox{exp}(-it\tilde{H}_{II})\Phi(0)=
\left(
  \begin{array}{ccc}
  1 &  &                                    \\
    & \mbox{cos}(gt)   & -i\mbox{sin}(gt)   \\
    & -i\mbox{sin}(gt) & \mbox{cos}(gt)    
  \end{array}
\right)\Phi(0).
\end{equation}
As a result the solution we are looking for is 
\[
\Psi(t)=\mbox{e}^{-itE_{1}}
\left(
  \begin{array}{ccc}
   \mbox{e}^{it\Delta_{1}}&   &                \\
    & 1 &                                      \\
    &   & \mbox{e}^{-i(\phi_{2}+\omega_{2}t)} 
  \end{array}
\right)
\left(
  \begin{array}{ccc}
  1 &  &                                    \\
    & \mbox{cos}(gt)   & -i\mbox{sin}(gt)   \\
    & -i\mbox{sin}(gt) & \mbox{cos}(gt)   
  \end{array}
\right)\Phi(0).
\]
However let us rewrite this. It is easy to see 
\[
\mbox{e}^{-itE_{1}}
\left(
  \begin{array}{ccc}
   \mbox{e}^{it\Delta_{1}}&   &                \\
    & 1 &                                      \\
    &   & \mbox{e}^{-i(\phi_{2}+\omega_{2}t)} 
  \end{array}
\right)
=
\mbox{e}^{-itE_{0}}
\left(
  \begin{array}{ccc}
   1 &   &                                                     \\
     & \mbox{e}^{-it\Delta_{1}} &                              \\
     &   & \mbox{e}^{-i\{\phi_{2}+(\omega_{2}+\Delta_{1})t\}} 
  \end{array}
\right)
\]
by the relation $\Delta_{1}=E_{1}-E_{0}$, so 
\begin{equation}
\label{eq:full-solution-2}
\Psi(t)=\mbox{e}^{-itE_{0}}
\left(
  \begin{array}{ccc}
   1&   &                                                     \\
    & \mbox{e}^{-it\Delta_{1}} &                              \\
    &   & \mbox{e}^{-i\{\phi_{2}+(\omega_{2}+\Delta_{1})t\}} 
  \end{array}
\right)
\left(
  \begin{array}{ccc}
  1 &  &                                    \\
    & \mbox{cos}(gt)   & -i\mbox{sin}(gt)   \\
    & -i\mbox{sin}(gt) & \mbox{cos}(gt)   
  \end{array}
\right)\Phi(0).
\end{equation}

For the latter use we set 
\begin{equation}
\label{eq:unitary-2}
U_{2}(t,0)=\mbox{e}^{-itE_{1}}
\left(
  \begin{array}{ccc}
   1 &   &                                                     \\
     & \mbox{e}^{-it\Delta_{1}} &                              \\
     &   & \mbox{e}^{-i\{\phi_{2}+(\omega_{2}+\Delta_{1})t\}} 
  \end{array}
\right)
\left(
  \begin{array}{ccc}
  1 &  &                                  \\
    & \mbox{cos}(gt) & -i\mbox{sin}(gt)   \\
    & -i\mbox{sin}(gt) & \mbox{cos}(gt)   
  \end{array}
\right). 
\end{equation}

\subsection{Unitary Transformation of type III}

We consider the following case. That is, we shoot the atom with a laser field 
having the frequency equal to the energy difference $E_{2}-E_{0}$. 
See the following picture.

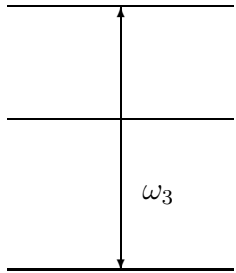
\vspace{5mm}
\begin{center}
\setlength{\unitlength}{1mm}   %
%
\end{center}
\vspace{-15mm}

The Hamiltonian that we are treating is 
\begin{equation}
\label{eq:hamiltonian-3}
H_{III}=
\left(
  \begin{array}{ccc}
   E_{0} &  & g\mbox{e}^{i(\phi_{3}+\omega_{3}t)}    \\
         & E_{1} &                                   \\
   g\mbox{e}^{-i(\phi_{3}+\omega_{3}t)}  &  & E_{2}
  \end{array}
\right)
=E_{0}{\bf 1}_{3}+
\left(
  \begin{array}{ccc}
    0 &  & g\mbox{e}^{i(\phi_{3}+\omega_{3}t)}            \\
      & \Delta_{1} &                                      \\
    g\mbox{e}^{-i(\phi_{3}+\omega_{3}t)} &  & \Delta_{2}
  \end{array}
\right).
\end{equation}

We would like to solve the Schr{\" o}dinger equation 
\begin{equation}
\label{eq:schrodinger-equation-3}
i\frac{d}{dt}\Psi=H_{III}\Psi.
\end{equation}

If we note the following decomposition 
\begin{eqnarray}
&&\left(
  \begin{array}{ccc}
   0  &  & g\mbox{e}^{i(\phi_{3}+\omega_{3}t)}            \\
      & \Delta_{1} &                                      \\
   g\mbox{e}^{-i(\phi_{3}+\omega_{3}t)} &  & \Delta_{2}
  \end{array}
\right)
=               \nonumber \\
&&\left(
  \begin{array}{ccc}
  1 &   &                                       \\
    & \mbox{e}^{-it\Delta_{1}} &                \\
    &   & \mbox{e}^{-i(\phi_{3}+\omega_{3}t)} 
  \end{array}
\right)
\left(
  \begin{array}{ccc}
    0 &  & g            \\
      & \Delta_{1} &    \\
    g &   & \Delta_{2}
  \end{array}
\right)
\left(
  \begin{array}{ccc}
  1 &   &                                      \\
    & \mbox{e}^{it\Delta_{1}} &                \\
    &   & \mbox{e}^{i(\phi_{3}+\omega_{3}t)} 
  \end{array}
\right)   
\end{eqnarray}
and define 
\begin{equation}
\Phi=\mbox{e}^{itE_{0}}
\left(
  \begin{array}{ccc}
  1 & &                                      \\
    & \mbox{e}^{it\Delta_{1}} &              \\
    & & \mbox{e}^{i(\phi_{3}+\omega_{3}t)} 
  \end{array}
\right)\Psi 
\end{equation}
, then it is not difficult to see 
\begin{equation}
i\frac{d}{dt}\Phi=
\left(
  \begin{array}{ccc}
   0 &   & g                      \\
     & 0 &                        \\
   g &   & \Delta_{2}-\omega_{3}
  \end{array}
\right)\Phi
\equiv 
\tilde{H}_{III}\Phi.
\end{equation}
Here we set the resonance condition 
\begin{equation}
\label{eq:resonance-3}
  \Delta_{2}=\omega_{3},
\end{equation}
so the solution is easily obtained to be 
\begin{equation}
\Phi(t)=\mbox{exp}(-it\tilde{H}_{III})\Phi(0)=
\left(
  \begin{array}{ccc}
   \mbox{cos}(gt)   &   & -i\mbox{sin}(gt)  \\
                    & 1 &                   \\
   -i\mbox{sin}(gt) &   & \mbox{cos}(gt) 
  \end{array}
\right)\Phi(0).
\end{equation}
As a result the solution we are looking for is 
\begin{equation}
\label{eq:full-solution-3}
\Psi(t)=\mbox{e}^{-itE_{0}}
\left(
  \begin{array}{ccc}
  1 & &                                       \\
    & \mbox{e}^{-it\Delta_{1}} &              \\
    & & \mbox{e}^{-i(\phi_{3}+\omega_{3}t)} 
  \end{array}
\right)
\left(
  \begin{array}{ccc}
   \mbox{cos}(gt)   &   & -i\mbox{sin}(gt)  \\
                    & 1 &                   \\
   -i\mbox{sin}(gt) &   & \mbox{cos}(gt) 
  \end{array}
\right)\Phi(0).
\end{equation}

For the latter use we set 
\begin{equation}
\label{eq:unitary-3}
U_{3}(t,0)=\mbox{e}^{-itE_{0}}
\left(
  \begin{array}{ccc}
  1 & &                                       \\
    & \mbox{e}^{-it\Delta_{1}} &              \\
    & & \mbox{e}^{-i(\phi_{3}+\omega_{3}t)} 
  \end{array}
\right)
\left(
  \begin{array}{ccc}
   \mbox{cos}(gt)   &   & -i\mbox{sin}(gt)  \\
                    & 1 &                   \\
   -i\mbox{sin}(gt) &   & \mbox{cos}(gt) 
  \end{array}
\right).
\end{equation}

\subsection{Unitary Transformation of type IV}

We consider the following case. That is, we shoot the atom with two laser 
fields having the frequencies equal to the energy differences $E_{1}-E_{0}$ 
and $E_{2}-E_{1}$ respectively. 
See the following picture.

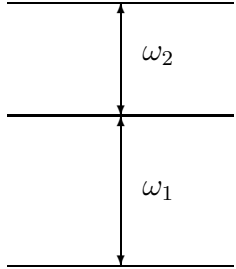
\vspace{5mm}
\begin{center}
\setlength{\unitlength}{1mm}   %
%
\end{center}
\vspace{-15mm}

The Hamiltonian that we are treating is 
\begin{eqnarray}
\label{eq:hamiltonian-4}
H_{IV}&=&
\left(
  \begin{array}{ccc}
   E_{0} & g\mbox{e}^{i(\phi_{1}+\omega_{1}t)} & 0    \\
   g\mbox{e}^{-i(\phi_{1}+\omega_{1}t)} & E_{1} & 
   g\mbox{e}^{i(\phi_{2}+\omega_{2}t)}                \\
   0 & g\mbox{e}^{-i(\phi_{2}+\omega_{2}t)} & E_{2}
  \end{array}
\right)        \nonumber \\
&=&
E_{0}{\bf 1}_{3}+
\left(
  \begin{array}{ccc}
   0 & g\mbox{e}^{i(\phi_{1}+\omega_{1}t)} & 0           \\
   g\mbox{e}^{-i(\phi_{1}+\omega_{1}t)} & \Delta_{1} & 
   g\mbox{e}^{i(\phi_{2}+\omega_{2}t)}                   \\
   0 & g\mbox{e}^{-i(\phi_{2}+\omega_{2}t)} & \Delta_{2}
  \end{array}
\right).
\end{eqnarray}

We would like to solve the Schr{\" o}dinger equation 
\begin{equation}
\label{eq:schrodinger-equation-4}
i\frac{d}{dt}\Psi=H_{IV}\Psi.
\end{equation}

If we note the following decomposition 
\begin{eqnarray}
&&\left(
  \begin{array}{ccc}
   0 & g\mbox{e}^{i(\phi_{1}+\omega_{1}t)} & 0           \\
   g\mbox{e}^{-i(\phi_{1}+\omega_{1}t)} & \Delta_{1} & 
   g\mbox{e}^{i(\phi_{2}+\omega_{2}t)}                   \\
   0 & g\mbox{e}^{-i(\phi_{2}+\omega_{2}t)} & \Delta_{2}
  \end{array}
\right)
=               \nonumber \\
&&\left(
  \begin{array}{ccc}
  1 & &                                            \\
    & \mbox{e}^{-i(\phi_{1}+\omega_{1}t)} &        \\
    & & \mbox{e}^{-i(\phi_{1}+\omega_{1}t+\phi_{2}+\omega_{2}t)}
  \end{array}
\right)
\left(
  \begin{array}{ccc}
   0 & g & 0           \\
   g & \Delta_{1} & g  \\
   0 & g & \Delta_{2}
  \end{array}
\right)
\left(
  \begin{array}{ccc}
  1 & &                                           \\
    & \mbox{e}^{i(\phi_{1}+\omega_{1}t)} &        \\
    & & \mbox{e}^{i(\phi_{1}+\omega_{1}t+\phi_{2}+\omega_{2}t)}
  \end{array}
\right)         \nonumber \\ 
&&{}  
\end{eqnarray}
and define 
\begin{equation}
\Phi=\mbox{e}^{itE_{0}}
\left(
  \begin{array}{ccc}
  1 & &                                           \\
    & \mbox{e}^{i(\phi_{1}+\omega_{1}t)} &        \\
    & & \mbox{e}^{i(\phi_{1}+\omega_{1}t+\phi_{2}+\omega_{2}t)}
  \end{array}
\right)\Psi 
\end{equation}
, then it is not difficult to see 
\begin{equation}
i\frac{d}{dt}\Phi=
\left(
  \begin{array}{ccc}
   0 & g & 0  \\
   g & \Delta_{1}-\omega_{1} & g \\
   0 & g & \Delta_{2}-\omega_{1}-\omega_{2}
  \end{array}
\right)\Phi
\equiv 
\tilde{H}_{IV}\Phi.
\end{equation}
Here we set the resonance condition 
\begin{equation}
\label{eq:resonance-4}
 \Delta_{1}=\omega_{1}\quad \mbox{and}\quad \Delta_{2}=\omega_{1}+\omega_{2}
 \Longleftrightarrow 
 \Delta_{1}=\omega_{1}\quad \mbox{and}\quad \Delta_{2}-\Delta_{1}=\omega_{2},
\end{equation}
so the solution is obtained to be 
\begin{eqnarray}
\Phi(t)&=&\mbox{exp}(-it\tilde{H}_{IV})\Phi(0)  \nonumber \\
&=&\frac{1}{2}
\left(
  \begin{array}{ccc}
     1+\mbox{cos}(\sqrt{2}gt)& -i\sqrt{2}\mbox{sin}(\sqrt{2}gt)& 
    -1+\mbox{cos}(\sqrt{2}gt)  \\
     -i\sqrt{2}\mbox{sin}(\sqrt{2}gt)& 2\mbox{cos}(\sqrt{2}gt)& 
     -i\sqrt{2}\mbox{sin}(\sqrt{2}gt)  \\
     -1+\mbox{cos}(\sqrt{2}gt)& -i\sqrt{2}\mbox{sin}(\sqrt{2}gt)&
      1+\mbox{cos}(\sqrt{2}gt)
  \end{array}
\right)\Phi(0). 
\end{eqnarray}
As a result the solution we are looking for is 
\begin{eqnarray}
\Psi(t)&=&\mbox{e}^{-itE_{0}}
\left(
  \begin{array}{ccc}
  1 & &                                                          \\
    & \mbox{e}^{-i(\phi_{1}+\omega_{1}t)} &                      \\
    & & \mbox{e}^{-i(\phi_{1}+\omega_{1}t+\phi_{2}+\omega_{2}t)}
  \end{array}
\right)\times        \nonumber \\
&&
\left(
  \begin{array}{ccc}
     \frac{1+\mbox{cos}(\sqrt{2}gt)}{2}& 
   -i\frac{\mbox{sin}(\sqrt{2}gt)}{\sqrt{2}}& 
     \frac{-1+\mbox{cos}(\sqrt{2}gt)}{2}              \\
   -i\frac{\mbox{sin}(\sqrt{2}gt)}{\sqrt{2}}& \mbox{cos}(\sqrt{2}gt)& 
   -i\frac{\mbox{sin}(\sqrt{2}gt)}{\sqrt{2}}          \\
     \frac{-1+\mbox{cos}(\sqrt{2}gt)}{2}& 
   -i\frac{\mbox{sin}(\sqrt{2}gt)}{\sqrt{2}}&
     \frac{1+\mbox{cos}(\sqrt{2}gt)}{2}
  \end{array}
\right)\Phi(0). 
\end{eqnarray}

For the latter use we set 
\begin{eqnarray}
\label{eq:unitary-4}
U_{4}(t,0)&=&\mbox{e}^{-itE_{0}}
\left(
  \begin{array}{ccc}
  1 & &                                                          \\
    & \mbox{e}^{-i(\phi_{1}+\omega_{1}t)} &                      \\
    & & \mbox{e}^{-i(\phi_{1}+\omega_{1}t+\phi_{2}+\omega_{2}t)}
  \end{array}
\right)\times        \nonumber \\
&&
\left(
  \begin{array}{ccc}
     \frac{1+\mbox{cos}(\sqrt{2}gt)}{2}& 
   -i\frac{\mbox{sin}(\sqrt{2}gt)}{\sqrt{2}}& 
     \frac{-1+\mbox{cos}(\sqrt{2}gt)}{2}              \\
   -i\frac{\mbox{sin}(\sqrt{2}gt)}{\sqrt{2}}& \mbox{cos}(\sqrt{2}gt)& 
   -i\frac{\mbox{sin}(\sqrt{2}gt)}{\sqrt{2}}          \\
     \frac{-1+\mbox{cos}(\sqrt{2}gt)}{2}& 
   -i\frac{\mbox{sin}(\sqrt{2}gt)}{\sqrt{2}}&
     \frac{1+\mbox{cos}(\sqrt{2}gt)}{2}
  \end{array}
\right). 
\end{eqnarray}

\subsection{Unitary Transformation of type V}

We consider the following case. That is, we shoot the atom with two laser 
fields having the frequencies equal to the energy differences $E_{1}-E_{0}$ 
and $E_{2}-E_{0}$ respectively. 
See the following picture.

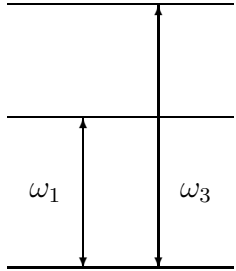
\vspace{5mm}
\begin{center}
\setlength{\unitlength}{1mm}   %
%
\end{center}
\vspace{-15mm}

The Hamiltonian that we are treating is 
\begin{eqnarray}
\label{eq:hamiltonian-5}
H_{V}&=&
\left(
  \begin{array}{ccc}
   E_{0} & g\mbox{e}^{i(\phi_{1}+\omega_{1}t)} & 
   g\mbox{e}^{i(\phi_{3}+\omega_{3}t)}                \\
   g\mbox{e}^{-i(\phi_{1}+\omega_{1}t)} & E_{1} & 0   \\
   g\mbox{e}^{-i(\phi_{3}+\omega_{3}t)}& 0 & E_{2}
  \end{array}
\right)           \nonumber \\
&=&
E_{0}{\bf 1}_{3}+
\left(
  \begin{array}{ccc}
   0 & g\mbox{e}^{i(\phi_{1}+\omega_{1}t)} & 
   g\mbox{e}^{i(\phi_{3}+\omega_{3}t)}                    \\
   g\mbox{e}^{-i(\phi_{1}+\omega_{1}t)} & \Delta_{1} & 0  \\
   g\mbox{e}^{-i(\phi_{3}+\omega_{3}t)}& 0 & \Delta_{2}
  \end{array}
\right).
\end{eqnarray}

We would like to solve the Schr{\" o}dinger equation 
\begin{equation}
\label{eq:schrodinger-equation-5}
i\frac{d}{dt}\Psi=H_{V}\Psi.
\end{equation}

If we note the following decomposition 
\begin{eqnarray}
&&\left(
  \begin{array}{ccc}
   0 & g\mbox{e}^{i(\phi_{1}+\omega_{1}t)} & 
   g\mbox{e}^{i(\phi_{3}+\omega_{3}t)}                     \\
   g\mbox{e}^{-i(\phi_{1}+\omega_{1}t)} & \Delta_{1} & 0   \\
   g\mbox{e}^{-i(\phi_{3}+\omega_{3}t)} & 0 & \Delta_{2}
  \end{array}
\right)
=               \nonumber \\
&&\left(
  \begin{array}{ccc}
  1 &                                        &     \\
    & \mbox{e}^{-i(\phi_{1}+\omega_{1}t)}    &     \\
    &  & \mbox{e}^{-i(\phi_{3}+\omega_{3}t)}
  \end{array}
\right)
\left(
  \begin{array}{ccc}
   0 & g & g             \\
   g & \Delta_{1} & 0    \\
   g & 0 & \Delta_{2}
  \end{array}
\right)
\left(
  \begin{array}{ccc}
  1 & &                                           \\
    & \mbox{e}^{i(\phi_{1}+\omega_{1}t)} &        \\
    & & \mbox{e}^{i(\phi_{3}+\omega_{3}t)}
  \end{array}
\right)         \nonumber \\ 
&&{}  
\end{eqnarray}
and define 
\begin{equation}
\Phi=\mbox{e}^{itE_{0}}
\left(
  \begin{array}{ccc}
  1 & &                                        \\
    & \mbox{e}^{i(\phi_{1}+\omega_{1}t)} &     \\
    & & \mbox{e}^{i(\phi_{3}+\omega_{3}t)}
  \end{array}
\right)\Psi 
\end{equation}
, then it is not difficult to see 
\begin{equation}
i\frac{d}{dt}\Phi=
\left(
  \begin{array}{ccc}
   0 & g & g                       \\
   g & \Delta_{1}-\omega_{1} & 0   \\
   g & 0 & \Delta_{2}-\omega_{3}
  \end{array}
\right)\Phi
\equiv 
\tilde{H}_{V}\Phi.
\end{equation}
Here we set the resonance condition 
\begin{equation}
\label{eq:resonance-5}
 \Delta_{1}=\omega_{1}\quad \mbox{and}\quad \Delta_{2}=\omega_{3},
\end{equation}
so the solution is obtained to be 
\begin{eqnarray}
\Phi(t)&=&\mbox{exp}(-it\tilde{H}_{V})\Phi(0)  \nonumber \\
&=&
\left(
  \begin{array}{ccc}
     \mbox{cos}(\sqrt{2}gt)& -i\frac{\mbox{sin}(\sqrt{2}gt)}{\sqrt{2}}& 
     -i\frac{\mbox{sin}(\sqrt{2}gt)}{\sqrt{2}}           \\
     -i\frac{\mbox{sin}(\sqrt{2}gt)}{\sqrt{2}}& 
     \frac{1+\mbox{cos}(\sqrt{2}gt)}{2}& 
     \frac{-1+\mbox{cos}(\sqrt{2}gt)}{2}                 \\
     -i\frac{\mbox{sin}(\sqrt{2}gt)}{\sqrt{2}}& 
     \frac{-1+\mbox{cos}(\sqrt{2}gt)}{2}& 
     \frac{1+\mbox{cos}(\sqrt{2}gt)}{2}       
  \end{array}
\right)\Phi(0). 
\end{eqnarray}
As a result the solution we are looking for is 
\begin{eqnarray}
\Psi(t)&=&\mbox{e}^{-itE_{0}}
\left(
  \begin{array}{ccc}
  1 & &                                        \\
    & \mbox{e}^{-i(\phi_{1}+\omega_{1}t)} &    \\
    & & \mbox{e}^{-i(\phi_{3}+\omega_{3}t)}
  \end{array}
\right)\times        \nonumber \\
&&
\left(
  \begin{array}{ccc}
     \mbox{cos}(\sqrt{2}gt)& -i\frac{\mbox{sin}(\sqrt{2}gt)}{\sqrt{2}}& 
     -i\frac{\mbox{sin}(\sqrt{2}gt)}{\sqrt{2}}           \\
     -i\frac{\mbox{sin}(\sqrt{2}gt)}{\sqrt{2}}& 
     \frac{1+\mbox{cos}(\sqrt{2}gt)}{2}& 
     \frac{-1+\mbox{cos}(\sqrt{2}gt)}{2}                 \\
     -i\frac{\mbox{sin}(\sqrt{2}gt)}{\sqrt{2}}& 
     \frac{-1+\mbox{cos}(\sqrt{2}gt)}{2}& 
     \frac{1+\mbox{cos}(\sqrt{2}gt)}{2}       
  \end{array}
\right)\Phi(0). 
\end{eqnarray}

For the latter use we set 
\begin{eqnarray}
\label{eq:unitary-5}
U_{5}(t,0)&=&\mbox{e}^{-itE_{0}}
\left(
  \begin{array}{ccc}
  1 & &                                        \\
    & \mbox{e}^{-i(\phi_{1}+\omega_{1}t)} &    \\
    & & \mbox{e}^{-i(\phi_{3}+\omega_{3}t)}
  \end{array}
\right)\times        \nonumber \\
&&
\left(
  \begin{array}{ccc}
     \mbox{cos}(\sqrt{2}gt)& -i\frac{\mbox{sin}(\sqrt{2}gt)}{\sqrt{2}}& 
     -i\frac{\mbox{sin}(\sqrt{2}gt)}{\sqrt{2}}           \\
     -i\frac{\mbox{sin}(\sqrt{2}gt)}{\sqrt{2}}& 
     \frac{1+\mbox{cos}(\sqrt{2}gt)}{2}& 
     \frac{-1+\mbox{cos}(\sqrt{2}gt)}{2}                 \\
     -i\frac{\mbox{sin}(\sqrt{2}gt)}{\sqrt{2}}& 
     \frac{-1+\mbox{cos}(\sqrt{2}gt)}{2}& 
     \frac{1+\mbox{cos}(\sqrt{2}gt)}{2}       
  \end{array}
\right). 
\end{eqnarray}

\subsection{Unitary Transformation of type VI}

We consider the following case. That is, we shoot the atom with two laser 
fields having the frequencies equal to the energy difference $E_{2}-E_{0}$ 
and $E_{2}-E_{1}$ respectively. 
See the following picture.

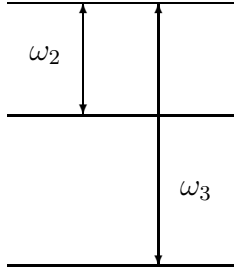
\vspace{5mm}
\begin{center}
\setlength{\unitlength}{1mm}   %
%
\end{center}
\vspace{-15mm}

The Hamiltonian that we are treating is 
\begin{eqnarray}
\label{eq:hamiltonian-6}
H_{VI}&=&
\left(
  \begin{array}{ccc}
   E_{0} & 0 & g\mbox{e}^{i(\phi_{3}+\omega_{3}t)}   \\
   0 & E_{1} & g\mbox{e}^{i(\phi_{2}+\omega_{2}t)}   \\
   g\mbox{e}^{-i(\phi_{3}+\omega_{3}t)} & 
   g\mbox{e}^{-i(\phi_{2}+\omega_{2}t)} & E_{2}
  \end{array}
\right)           \nonumber \\
&=&E_{0}{\bf 1}_{3}+
\left(
  \begin{array}{ccc}
   0 & 0 & g\mbox{e}^{i(\phi_{3}+\omega_{3}t)}            \\
   0 & \Delta_{1} & g\mbox{e}^{i(\phi_{2}+\omega_{2}t)}   \\
   g\mbox{e}^{-i(\phi_{3}+\omega_{3}t)} & 
   g\mbox{e}^{-i(\phi_{2}+\omega_{2}t)} & \Delta_{2}
  \end{array}
\right).
\end{eqnarray}

We would like to solve the Schr{\" o}dinger equation 
\begin{equation}
\label{eq:schrodinger-equation-6}
i\frac{d}{dt}\Psi=H_{VI}\Psi.
\end{equation}

If we note the following decomposition 
\begin{eqnarray}
&&\left(
  \begin{array}{ccc}
   0 & 0 & g\mbox{e}^{i(\phi_{3}+\omega_{3}t)}            \\
   0 & \Delta_{1} & g\mbox{e}^{i(\phi_{2}+\omega_{2}t)}   \\
   g\mbox{e}^{-i(\phi_{3}+\omega_{3}t)} & 
   g\mbox{e}^{-i(\phi_{2}+\omega_{2}t)} & \Delta_{2}
  \end{array}
\right)
=               \nonumber \\
&&\left(
  \begin{array}{ccc} 
  1 & &                                                           \\
    & \mbox{e}^{-i(\phi_{3}+\omega_{3}t-\phi_{2}-\omega_{2}t)} &  \\
    & & \mbox{e}^{-i(\phi_{3}+\omega_{3}t)}
  \end{array}
\right)
\left(
  \begin{array}{ccc}
   0 & 0 & g             \\
   0 & \Delta_{1} & g    \\
   g & g & \Delta_{2}
  \end{array}
\right)
\left(
  \begin{array}{ccc}
  1 & &                                                          \\
    & \mbox{e}^{i(\phi_{3}+\omega_{3}t-\phi_{2}-\omega_{2}t)} &  \\
    & & \mbox{e}^{i(\phi_{3}+\omega_{3}t)} 
  \end{array}
\right)         \nonumber \\ 
&&{}  
\end{eqnarray}
and define 
\begin{equation}
\Phi=\mbox{e}^{itE_{0}}
\left(
  \begin{array}{ccc}
  1 & &                                                           \\
    & \mbox{e}^{i(\phi_{3}+\omega_{3}t-\phi_{2}-\omega_{2}t)} &   \\
    & & \mbox{e}^{i(\phi_{3}+\omega_{3}t)}
  \end{array}
\right)\Psi 
\end{equation}
, then it is not difficult to see 
\begin{equation}
i\frac{d}{dt}\Phi=
\left(
  \begin{array}{ccc}
   0 & 0 & g                                  \\
   0 & \Delta_{1}-\omega_{3}+ \omega_{2} & g  \\
   g & g & \Delta_{2}-\omega_{3}
  \end{array}
\right)\Phi
\equiv 
\tilde{H}_{VI}\Phi.
\end{equation}
Here we set the resonance condition 
\begin{equation}
\label{eq:resonance-6}
 \Delta_{1}=\omega_{3}-\omega_{2}\quad \mbox{and}\quad \Delta_{2}=\omega_{3}
 \Longleftrightarrow 
 \Delta_{2}-\Delta_{1}=\omega_{2}\quad \mbox{and}\quad \Delta_{2}=\omega_{3},
\end{equation}
so the solution is obtained to be 
\begin{eqnarray}
\Phi(t)&=&\mbox{exp}(-it\tilde{H}_{VI})\Phi(0)  \nonumber \\
&=&
\left(
  \begin{array}{ccc}
     \frac{1+\mbox{cos}(\sqrt{2}gt)}{2}& 
     \frac{-1+\mbox{cos}(\sqrt{2}gt)}{2}& 
   -i\frac{\mbox{sin}(\sqrt{2}gt)}{\sqrt{2}}       \\
     \frac{-1+\mbox{cos}(\sqrt{2}gt)}{2}& 
     \frac{1+\mbox{cos}(\sqrt{2}gt)}{2}& 
   -i\frac{\mbox{sin}(\sqrt{2}gt)}{\sqrt{2}}       \\
   -i\frac{\mbox{sin}(\sqrt{2}gt)}{\sqrt{2}}& 
   -i\frac{\mbox{sin}(\sqrt{2}gt)}{\sqrt{2}}&
     \mbox{cos}(\sqrt{2}gt)
  \end{array}
\right)\Phi(0). 
\end{eqnarray}

As a result the solution we are looking for is 
\begin{eqnarray}
\Psi(t)&=&\mbox{e}^{-itE_{0}}
\left(
  \begin{array}{ccc}
  1 & &                                                           \\
    & \mbox{e}^{i(\phi_{3}+\omega_{3}t-\phi_{2}-\omega_{2}t)} &   \\
    & & \mbox{e}^{i(\phi_{3}+\omega_{3}t)}
  \end{array}
\right)\times       \nonumber \\
&&
\left(
  \begin{array}{ccc}
     \frac{1+\mbox{cos}(\sqrt{2}gt)}{2}& 
     \frac{-1+\mbox{cos}(\sqrt{2}gt)}{2}& 
   -i\frac{\mbox{sin}(\sqrt{2}gt)}{\sqrt{2}}       \\
     \frac{-1+\mbox{cos}(\sqrt{2}gt)}{2}& 
     \frac{1+\mbox{cos}(\sqrt{2}gt)}{2}& 
   -i\frac{\mbox{sin}(\sqrt{2}gt)}{\sqrt{2}}       \\
   -i\frac{\mbox{sin}(\sqrt{2}gt)}{\sqrt{2}}& 
   -i\frac{\mbox{sin}(\sqrt{2}gt)}{\sqrt{2}}&
     \mbox{cos}(\sqrt{2}gt)
  \end{array}
\right)\Phi(0).
\end{eqnarray}

For the latter use we set 
\begin{eqnarray}
\label{eq:unitary-6}
U_{6}(t,0)&=&\mbox{e}^{-itE_{0}}
\left(
  \begin{array}{ccc}
  1 & &                                                           \\
    & \mbox{e}^{i(\phi_{3}+\omega_{3}t-\phi_{2}-\omega_{2}t)} &   \\
    & & \mbox{e}^{i(\phi_{3}+\omega_{3}t)}
  \end{array}
\right)\times       \nonumber \\
&&
\left(
  \begin{array}{ccc}
     \frac{1+\mbox{cos}(\sqrt{2}gt)}{2}& 
     \frac{-1+\mbox{cos}(\sqrt{2}gt)}{2}& 
   -i\frac{\mbox{sin}(\sqrt{2}gt)}{\sqrt{2}}       \\
     \frac{-1+\mbox{cos}(\sqrt{2}gt)}{2}& 
     \frac{1+\mbox{cos}(\sqrt{2}gt)}{2}& 
   -i\frac{\mbox{sin}(\sqrt{2}gt)}{\sqrt{2}}       \\
   -i\frac{\mbox{sin}(\sqrt{2}gt)}{\sqrt{2}}& 
   -i\frac{\mbox{sin}(\sqrt{2}gt)}{\sqrt{2}}&
     \mbox{cos}(\sqrt{2}gt)
  \end{array}
\right).
\end{eqnarray}

\subsection{Unitary Transformation of type VII}

We consider the following case. That is, we shoot the atom with three laser 
fields having the frequencies equal to the energy difference $E_{1}-E_{0}$, 
$E_{2}-E_{1}$ and $E_{2}-E_{0}$ respectively. 
See the following picture.

\vspace{5mm}
\begin{center}
\setlength{\unitlength}{1mm}   %
%
\end{center}
\vspace{-15mm}

The Hamiltonian is 
\begin{eqnarray}
\label{eq:hamiltonian-7}
H_{VII}&=&
\left(
  \begin{array}{ccc}
   E_{0} & g\mbox{e}^{i(\phi_{1}+\omega_{1}t)} & 
   g\mbox{e}^{i(\phi_{3}+\omega_{3}t)}               \\
   g\mbox{e}^{-i(\phi_{1}+\omega_{1}t)} & E_{1} & 
   g\mbox{e}^{i(\phi_{2}+\omega_{2}t)}               \\
   g\mbox{e}^{-i(\phi_{3}+\omega_{3}t)} & 
   g\mbox{e}^{-i(\phi_{2}+\omega_{2}t)} & E_{2}
  \end{array}
\right)          \nonumber \\
&=&
E_{0}{\bf 1}_{3}+
\left(
  \begin{array}{ccc}
   0 & g\mbox{e}^{i(\phi_{1}+\omega_{1}t)} & 
   g\mbox{e}^{i(\phi_{3}+\omega_{3}t)}                   \\
   g\mbox{e}^{-i(\phi_{1}+\omega_{1}t)} & \Delta_{1} & 
   g\mbox{e}^{i(\phi_{2}+\omega_{2}t)}                   \\
   g\mbox{e}^{-i(\phi_{3}+\omega_{3}t)} & 
   g\mbox{e}^{-i(\phi_{2}+\omega_{2}t)} & \Delta_{2}
  \end{array}
\right).
\end{eqnarray}
Here we assume the following condition 
\[
\omega_{3}=\omega_{1}+\omega_{2},
\]
which is a kind of consistency condition.  We note that without this 
assumption it is very difficult (almost impossible) to solve the following 
equation, see \cite{FHKW}. 

We would like to solve the Schr{\" o}dinger equation 
\begin{equation}
\label{eq:schrodinger-equation-7}
i\frac{d}{dt}\Psi=H_{VII}\Psi.
\end{equation}

If we note the following decomposition 
\begin{eqnarray}
&&\left(
  \begin{array}{ccc}
   0 & g\mbox{e}^{i(\phi_{1}+\omega_{1}t)} & 
   g\mbox{e}^{i(\phi_{3}+\omega_{3}t)}                  \\
   g\mbox{e}^{-i(\phi_{1}+\omega_{1}t)} & \Delta_{1} & 
   g\mbox{e}^{i(\phi_{2}+\omega_{2}t)}                  \\
   g\mbox{e}^{-i(\phi_{3}+\omega_{3}t)} & 
   g\mbox{e}^{-i(\phi_{2}+\omega_{2}t)} & \Delta_{2}
  \end{array}
\right)
=               \nonumber \\
&&\left(
  \begin{array}{ccc} 
  1 & &                                                           \\
    & \mbox{e}^{-i(\phi_{1}+\omega_{1}t)} &                       \\
    & & \mbox{e}^{-i(\phi_{1}+\omega_{1}t+\phi_{2}+\omega_{2}t)}
  \end{array}
\right)
\left(
  \begin{array}{ccc}
   0 & g & g\mbox{e}^{i(\phi_{3}-\phi_{1}-\phi_{2})}            \\
   g & \Delta_{1} & g                                           \\
   g\mbox{e}^{-i(\phi_{3}-\phi_{1}-\phi_{2})} & g & \Delta_{2}
  \end{array}
\right) \times     \nonumber \\
&&
\left(
  \begin{array}{ccc}
  1 & &                                                          \\
    & \mbox{e}^{i(\phi_{1}+\omega_{1}t)} &                       \\
    & & \mbox{e}^{i(\phi_{1}+\omega_{1}t+\phi_{2}+\omega_{2}t)} 
  \end{array}
\right)  
\end{eqnarray}
and define 
\begin{equation}
\Phi=\mbox{e}^{itE_{0}}
\left(
  \begin{array}{ccc}
  1 & &                                                           \\
    & \mbox{e}^{i(\phi_{1}+\omega_{1}t)} &                        \\
    & & \mbox{e}^{i(\phi_{1}+\omega_{1}t+\phi_{2}+\omega_{2}t)}
  \end{array}
\right)\Psi 
\end{equation}
, then it is not difficult to see 
\begin{equation}
i\frac{d}{dt}\Phi=
\left(
  \begin{array}{ccc}
   0 & g & g\mbox{e}^{i(\phi_{3}-\phi_{1}-\phi_{2})}      \\
   g & \Delta_{1}-\omega_{1} & g                          \\
   g\mbox{e}^{-i(\phi_{3}-\phi_{1}-\phi_{2})} & g & 
   \Delta_{2}-\omega_{1}-\omega_{2}
  \end{array}
\right)\Phi
\equiv 
\tilde{H}_{VII}\Phi.
\end{equation}
Here we set the resonance condition 
\begin{equation}
\label{eq:resonance-7}
 \Delta_{1}=\omega_{1}\quad \mbox{and}\quad 
 \Delta_{2}=\omega_{1}+\omega_{2}=\omega_{3},
\end{equation}
and moreover the phase condition 
\begin{equation}
\label{eq:phase-7}
 \phi_{3}=\phi_{1}+\phi_{2},
\end{equation}
so the solution is obtained to be 
\begin{eqnarray}
\Phi(t)&=&\mbox{exp}(-it\tilde{H}_{VII})\Phi(0)  \nonumber \\
&=&\mbox{e}^{igt}
\left(
  \begin{array}{ccc}
     \frac{2+\mbox{exp}(-i3gt)}{3}&  \frac{-1+\mbox{exp}(-i3gt)}{3}& 
     \frac{-1+\mbox{exp}(-i3gt)}{3}                                    \\
     \frac{-1+\mbox{exp}(-i3gt)}{3}& \frac{2+\mbox{exp}(-i3gt)}{3}& 
     \frac{-1+\mbox{exp}(-i3gt)}{3}                                    \\
     \frac{-1+\mbox{exp}(-i3gt)}{3}& \frac{-1+\mbox{exp}(-i3gt)}{3}&
     \frac{2+\mbox{exp}(-i3gt)}{3}
  \end{array}
\right)\Phi(0). 
\end{eqnarray}

As a result the solution we are looking for is 
\begin{eqnarray}
\Psi(t)&=&\mbox{e}^{-itE_{0}}
\left(
  \begin{array}{ccc}
  1 & &                                                           \\
    & \mbox{e}^{-i(\phi_{1}+\omega_{1}t)}  &                      \\
    & & \mbox{e}^{-i(\phi_{1}+\omega_{1}t+\phi_{2}+\omega_{2}t)}
  \end{array}
\right)\times       \nonumber \\
&&\mbox{e}^{igt}
\left(
  \begin{array}{ccc}
     \frac{2+\mbox{exp}(-i3gt)}{3}&  \frac{-1+\mbox{exp}(-i3gt)}{3}& 
     \frac{-1+\mbox{exp}(-i3gt)}{3}                                    \\
     \frac{-1+\mbox{exp}(-i3gt)}{3}& \frac{2+\mbox{exp}(-i3gt)}{3}& 
     \frac{-1+\mbox{exp}(-i3gt)}{3}                                    \\
     \frac{-1+\mbox{exp}(-i3gt)}{3}& \frac{-1+\mbox{exp}(-i3gt)}{3}&
     \frac{2+\mbox{exp}(-i3gt)}{3}
  \end{array}
\right)\Phi(0). 
\end{eqnarray}

For the latter use we set 
\begin{eqnarray}
\label{eq:unitary-7}
U_{7}(t,0)&=&\mbox{e}^{-itE_{0}}
\left(
  \begin{array}{ccc}
  1 & &                                                           \\
    & \mbox{e}^{-i(\phi_{1}+\omega_{1}t)}  &                      \\
    & & \mbox{e}^{-i(\phi_{1}+\omega_{1}t+\phi_{2}+\omega_{2}t)}
  \end{array}
\right)\times       \nonumber \\
&&\mbox{e}^{igt}
\left(
  \begin{array}{ccc}
     \frac{2+\mbox{exp}(-i3gt)}{3}&  \frac{-1+\mbox{exp}(-i3gt)}{3}& 
     \frac{-1+\mbox{exp}(-i3gt)}{3}                                    \\
     \frac{-1+\mbox{exp}(-i3gt)}{3}& \frac{2+\mbox{exp}(-i3gt)}{3}& 
     \frac{-1+\mbox{exp}(-i3gt)}{3}                                    \\
     \frac{-1+\mbox{exp}(-i3gt)}{3}& \frac{-1+\mbox{exp}(-i3gt)}{3}&
     \frac{2+\mbox{exp}(-i3gt)}{3}
  \end{array}
\right). 
\end{eqnarray}

{\bf A comment is in order}. We have prepared unitary transformations of 
eight types constructed from Rabi oscillations. In the following section 
by making use of these ones we will construct several unitary matrices 
indispensable in Quantum Computation based on three level systems.

\section{N Level Systems $\cdots$ Basic Theory}

In this section we introduce a generalization of Pauli matrices in section 2 
which has been used in several situations in both Quantum Field Theory and 
Quantum Computation, and also introduce a generalized Walsh--Hadamard matrix 
that plays a crucial role in Quantum Computation based on $n$ level systems. 
See for example \cite{KF1}, Appendix B. 

\par \noindent
We construct them in terms of Rabi oscillations made in the preceding section 
in the three level systems.

\subsection{Basic Theory}

First of all we summarize the properties of Pauli matrices. 
By (\ref{eq:pauli}) $\sigma_{2}=i\sigma_{1}\sigma_{3}$, so that the essential 
elements of Pauli matrices are $\{\sigma_{1}, \sigma_{3}\}$ and they satisfy
\begin{equation}
\sigma_{1}^{2}=\sigma_{3}^{2}={\bf 1}_{2};\quad 
\sigma_{1}^{\dagger}=\sigma_{1},\
\sigma_{3}^{\dagger}=\sigma_{3};\quad 
\sigma_{3}\sigma_{1}=-\sigma_{1}\sigma_{3}=\mbox{e}^{i\pi}\sigma_{1}\sigma_{3}.
\end{equation}

The Walsh--Hadamard matrix is defined by 
\begin{equation}
   \label{eq:w-a}
   W = \frac{1}{\sqrt{2}}
     \left(
        \begin{array}{rr}
            1& 1 \\
            1& -1
        \end{array}
     \right)\ \in \ O(2)\ \subset U(2).
\end{equation}
This matrix (or transformation) is unitary and it plays a very important role 
in Quantum Computation. Moreover it is easy to realize it in Quantum Optics 
as shown in section 2. 
Let us list some important properties of $W$ :
\begin{eqnarray}
\label{eq:properties of W-H (1)}
      &&W^{2}={\bf 1}_{2},\ \ W^{\dagger}=W=W^{-1}, \\
\label{eq:properties of W-H (2)}
      &&\sigma_{1}= W\sigma_{3}W^{-1},
\end{eqnarray}
The check is very easy. 

\par \noindent
Let $\{\Sigma_{1}, \Sigma_{3}\}$ be the following matrices in $M(n,\fukuso)$
\begin{equation}
\label{eq:gener-pauli}
\Sigma_{1}=
\left(
\begin{array}{cccccc}
0&  &  &      &      &       1   \\
1& 0&  &      &      &           \\
  & 1& 0&      &      &          \\
  &  & 1& \cdot&      &          \\
  &  &  & \cdot& \cdot&          \\
  &  &  &      &    1 & 0
\end{array}
\right),      \qquad
\Sigma_{3}=
\left(
\begin{array}{cccccc}  
1&        &           &      &      &                  \\
  & \sigma&           &      &      &                  \\
  &       & {\sigma}^2&      &      &                  \\
  &       &           & \cdot&      &                  \\
  &       &           &      & \cdot&                  \\
  &       &           &      &      &  {\sigma}^{n-1}
\end{array}
\right)
\end{equation}
where $\sigma$ is a primitive root of unity ${\sigma}^{n}=1$ (
$\sigma=\mbox{e}^{\frac{2\pi i}{n}}$). We note that
\[
\bar{\sigma}=\sigma^{n-1},\quad
1+\sigma+\cdots+\sigma^{n-1}=0 .
\]
The two matrices
$\{\Sigma_{1}, \Sigma_{3}\}$ are generalizations of Pauli matrices
$\{\sigma_{1}, \sigma_{3}\}$, but they are not hermitian.
Here we list some of their important properties:
\begin{equation}
\Sigma_{1}^{n}=\Sigma_{3}^{n}={\bf 1}_{n}\quad ; \quad
\Sigma_{1}^{\dagger}=\Sigma_{1}^{n-1},\
\Sigma_{3}^{\dagger}=\Sigma_{3}^{n-1}\quad ; \quad
\Sigma_{3}\Sigma_{1}=\sigma \Sigma_{1}\Sigma_{3}\ .
\end{equation}
If we define a Vandermonde matrix $W$ based on $\sigma$ as
\begin{eqnarray}
\label{eq:Large-double}
W&=&\frac{1}{\sqrt{n}}
\left(
\begin{array}{ccccccc}
1&        1&     1&   \cdot & \cdot  & \cdot & 1             \\
1& \sigma^{n-1}& \sigma^{2(n-1)}&  \cdot& \cdot& \cdot& \sigma^{(n-1)^2} \\
1& \sigma^{n-2}& \sigma^{2(n-2)}&  \cdot& \cdot& \cdot& \sigma^{(n-1)(n-2)} \\
\cdot&  \cdot &  \cdot  &     &      &      & \cdot  \\
\cdot&  \cdot  & \cdot &      &      &      &  \cdot  \\
1& \sigma^{2}& \sigma^{4}& \cdot & \cdot & \cdot & \sigma^{2(n-1)} \\
1& \sigma & \sigma^{2}& \cdot& \cdot& \cdot& \sigma^{n-1}
\end{array}
\right), \\
\label{eq:Large-double-dagger}
W^{\dagger}&=&\frac{1}{\sqrt{n}}
\left(
\begin{array}{ccccccc}
1&        1&     1&   \cdot & \cdot  & \cdot & 1             \\
1& \sigma& \sigma^{2}&  \cdot& \cdot& \cdot& \sigma^{n-1} \\
1& \sigma^{2}& \sigma^{4}&  \cdot& \cdot& \cdot& \sigma^{2(n-1)} \\
\cdot&  \cdot &  \cdot  &     &      &      & \cdot  \\
\cdot&  \cdot  & \cdot &      &      &      &  \cdot  \\
1& \sigma^{n-2}& \sigma^{2(n-2)}& \cdot& \cdot& \cdot& \sigma^{(n-1)(n-2)} \\
1&    \sigma^{n-1} & \sigma^{2(n-1)}& \cdot& \cdot& \cdot& \sigma^{(n-1)^2}
\end{array}
\right),
\end{eqnarray}
then it is not difficult to see
\begin{equation}
\label{eq:diagonalization}
\Sigma_{1}=W\Sigma_{3}W^{\dagger}=W\Sigma_{3}W^{-1}.
\end{equation}
For example, for $n=3$
\begin{equation}
\Sigma_{1}=
\left(
\begin{array}{ccc}
0&  0& 1 \\
1&  0& 0 \\
0&  1& 0
\end{array}
\right), \quad 
\Sigma_{3}=
\left(
\begin{array}{ccc}
 1&       &            \\
  & \sigma&            \\
  &       & {\sigma}^2
\end{array}
\right), \quad 
W=\frac{1}{\sqrt{3}}
\left(
\begin{array}{ccc}
1&        1&     1         \\
1& \sigma^{2} & \sigma     \\
1& \sigma     & \sigma^{2}
\end{array}
\right) 
\end{equation}
where $\sigma=\mbox{e}^{\frac{2\pi i}{3}}=\frac{1}{2}(-1+i\sqrt{3})$ and 
\begin{eqnarray}
W\Sigma_{3}W^{\dagger}&=&\frac{1}{3}
\left(
\begin{array}{ccc}
1&          1&     1     \\
1& \sigma^{2}& \sigma    \\
1& \sigma    & \sigma^{2}
\end{array}
\right)
\left(
\begin{array}{ccc}
 1&       &            \\
  & \sigma&            \\
  &       & {\sigma}^2
\end{array}
\right)
\left(
\begin{array}{ccc}
1&        1&     1     \\
1& \sigma & \sigma^{2} \\
1& \sigma^{2}  & \sigma
\end{array}
\right)   \nonumber \\
&=&\frac{1}{3}
\left(
\begin{array}{ccc}
1& \sigma& \sigma^{2} \\
1&      1&  1         \\
1& \sigma^{2}& \sigma
\end{array}
\right)
\left(
\begin{array}{ccc}
1&        1&     1      \\
1& \sigma& \sigma^{2}   \\
1& \sigma^{2}  & \sigma
\end{array}
\right)
=\frac{1}{3}
\left(
\begin{array}{ccc}
0&  0& 3\\
3&  0& 0\\
0&  3& 0
\end{array}
\right)
=\Sigma_{1},  \nonumber
\end{eqnarray}
where we have used that $\sigma^{3}=1,\ \bar{\sigma}=\sigma^{2}\ \mbox{and}\
1+\sigma+\sigma^{2}=0$.

\par \noindent
That is, $\Sigma_{1}$ can be diagonalized by making use of $W$.

\par \noindent
{\bf A comment is in order}. Since $W$ corresponds to the Walsh--Hadamard 
matrix (\ref{eq:w-a}), so it may be possible to call $W$ the generalized 
Walsh--Hadamard matrix.

Now we define a matrix 
\begin{equation}
\label{eq:matrix-k}
K=
\left(
\begin{array}{ccccccc}
 1&   &   &      &      &    &     \\
  &   &   &      &      &    & 1   \\
  &   &   &      &      &  1 &     \\
  &   &   &      & \cdot&    &     \\
  &   &   & \cdot&      &    &     \\
  &   & 1 &      &      &    &     \\
  & 1 &   &      &      &    &   
\end{array}
\right)\ \in \ M(n,\fukuso)
\end{equation}
This matrix plays an important role in constructing the exchange gate 
in qudit theory, see \cite{KF4}. However when $n=2$ this becomes just the 
identity. 

Let us list some important properties of $W$ corresponding to 
(\ref{eq:properties of W-H (1)}), (\ref{eq:properties of W-H (2)}) :
\begin{eqnarray}
\label{eq:properties of Ge-W-H (1)}
      &&W^{2}=K,\ \ W^{\dagger}=KW=W^{-1}, \\
\label{eq:properties of Ge-W-H (2)}
      &&\Sigma_{1}= W\Sigma_{3}W^{-1},
\end{eqnarray}
The check is easy.

\subsection{Quantum Logic Gates on Three Level Systems}

From here we shall restrict to the case of $n=3$ and construct both 
$\{\Sigma_{1},\ \Sigma_{3}\}$ and $W$. 
Let us make a comment once more that we will change a time $t$ and phases 
$\phi_{j}$ as free parameters (we don't change the coupling constant $g$). 

Noting the fact 
\begin{equation}
\Sigma_{1}=
\left(
\begin{array}{ccc}
0&  0& 1 \\
1&  0& 0 \\
0&  1& 0
\end{array}
\right)
=
\left(
\begin{array}{ccc}
0&  0& 1 \\
0&  1& 0 \\
1&  0& 0
\end{array}
\right)
\left(
\begin{array}{ccc}
0&  1& 0 \\
1&  0& 0 \\
0&  0& 1
\end{array}
\right)
\end{equation}
we have only to construct each component. First let us construct the matrix 
\[
\left(
\begin{array}{ccc}
0&  1& 0 \\
1&  0& 0 \\
0&  0& 1
\end{array}
\right).
\]

From (\ref{eq:unitary-0}) we choose $t_{a}$ as $\Delta_{1}t_{a}=(3/2)\pi$ 
such as 
\[
U_{0}(t_{a},0)
=
\mbox{e}^{-iE_{0}t_{a}}
\left(
  \begin{array}{ccc}
  1 &                              &                               \\
    & \mbox{e}^{-i\Delta_{1}t_{a}} &                               \\
    &                              & \mbox{e}^{-i\Delta_{2}t_{a}}
  \end{array}
\right)
=
\mbox{e}^{-iE_{0}t_{a}}
\left(
  \begin{array}{ccc}
  1 &   &                            \\
    & i &                            \\
    &   & \mbox{e}^{-i\Delta_{2}t_{a}}
  \end{array}
\right).
\]
Next from (\ref{eq:unitary-1}) we choose $t_{b}\ (> t_{a})$ as 
$g(t_{b}-t_{a})=\pi/2$ such as 
\[
U_{1}(t_{b},t_{a})
=\mbox{e}^{-iE_{0}(t_{b}-t_{a})}
\left(
  \begin{array}{ccc}
  1 & &                                                   \\
    & \mbox{e}^{-i\{\omega_{1}(t_{b}-t_{a})+\phi_{1}\}} &   \\
    & & \mbox{e}^{-i\Delta_{2}(t_{b}-t_{a})}
  \end{array}
\right)
\left(
  \begin{array}{ccc}
      & -i &     \\
   -i &    &     \\
      &    &  1
  \end{array}
\right).
\]
We choose $t_{c}\ (> t_{b})$ as $\Delta_{1}(t_{c}-t_{b})=(3/2)\pi$ 
again in (\ref{eq:unitary-0}) such as 
\[
U_{0}(t_{c},t_{b})
=
\mbox{e}^{-iE_{0}(t_{c}-t_{b})}
\left(
  \begin{array}{ccc}
  1 &   &                                        \\
    & i &                                        \\
    &   & \mbox{e}^{-i\Delta_{2}(t_{c}-t_{b})}
  \end{array}
\right).
\]
From (\ref{eq:unitary-3}) we choose $t_{d}\ (> t_{c})$ as 
$g(t_{d}-t_{c})=2\pi$ such as 
\[
U_{3}(t_{d},t_{c})
=
\mbox{e}^{-iE_{0}(t_{d}-t_{c})}
\left(
  \begin{array}{ccc}
  1 &   &                                                 \\
    & \mbox{e}^{-i\Delta_{1}(t_{d}-t_{c})} &              \\
    &   & \mbox{e}^{-i\{\omega_{3}(t_{d}-t_{c})+\phi_{3}\}}
  \end{array}
\right).
\]
Therefore we have 
\begin{eqnarray}
&&U_{3}(t_{d},t_{c})U_{0}(t_{c},t_{b})U_{1}(t_{b},t_{a})U_{0}(t_{a},0)
      \nonumber \\
=&&
\mbox{e}^{-iE_{0}t_{d}}
\left(
  \begin{array}{ccc}
  1 &   &                                                            \\
    & \mbox{e}^{-i\{\omega_{1}(t_{b}-t_{a})+\Delta_{1}(t_{d}-t_{c})
    +\phi_{1}\}} &                                                   \\
    &   & \mbox{e}^{-i\{\omega_{3}(t_{d}-t_{c})+\Delta_{2}t_{c}+\phi_{3}\}}
  \end{array}
\right)
\left(
\begin{array}{ccc}
0&  1& 0 \\
1&  0& 0 \\
0&  0& 1
\end{array}
\right).           \nonumber 
\end{eqnarray}
In this case we cannot remove the phase $\mbox{e}^{-iE_{0}t_{d}}$. 
Here we again multiply the above matrix by the matrix 
\[
U_{0}(t_{e},t_{d})
=
\mbox{e}^{-iE_{0}(t_{e}-t_{d})}
\left(
  \begin{array}{ccc}
  1 &                     &                                        \\
    & \mbox{e}^{-i\Delta_{1}(t_{e}-t_{d})} &                       \\
    &                     & \mbox{e}^{-i\Delta_{2}(t_{e}-t_{d})}
  \end{array}
\right)
\]
and obtain 
\begin{eqnarray}
&&U_{0}(t_{e},t_{d})U_{3}(t_{d},t_{c})U_{0}(t_{c},t_{b})U_{1}(t_{b},t_{a})
U_{0}(t_{a},0)         \nonumber \\
=&&
\mbox{e}^{-iE_{0}t_{e}}
\left(
  \begin{array}{ccc}
  1 &   &                                                            \\
    & \mbox{e}^{-i\{\omega_{1}(t_{b}-t_{a})+\Delta_{1}(t_{e}-t_{c})
    +\phi_{1}\}} &                                                   \\
    &   & \mbox{e}^{-i\{\omega_{3}(t_{d}-t_{c})+
    \Delta_{2}(t_{e}-t_{d}+t_{c})+\phi_{3}\}}
  \end{array}
\right)
\left(
\begin{array}{ccc}
0&  1& 0 \\
1&  0& 0 \\
0&  0& 1
\end{array}
\right)            \nonumber 
\end{eqnarray}
with 
\[
\Delta_{1}t_{a}=(3/2)\pi,\ \ 
g(t_{b}-t_{a})=\pi/2,\ \ \Delta_{1}(t_{c}-t_{b})=(3/2)\pi,\ \ 
g(t_{d}-t_{c})=2\pi.
\]

If we choose $t_{e}\ (> t_{d})$ as 
\[
E_{0}t_{e}=2\pi k \quad \mbox{for some}\ k\ \in \futon
\]
and the phases $\phi_{1},\ \phi_{3}$ as 
\[
\mbox{e}^{-i\{\omega_{1}(t_{b}-t_{a})+\Delta_{1}(t_{e}-t_{c})+\phi_{1}\}}
=1,\quad 
\mbox{e}^{-i\{\omega_{3}(t_{d}-t_{c})+\Delta_{2}(t_{e}-t_{d}+t_{c})+
\phi_{3}\}}=1, 
\]
then we finally obtain 
\begin{equation}
U_{0}(t_{e},t_{d})U_{3}(t_{d},t_{c})U_{0}(t_{c},t_{b})U_{1}(t_{b},t_{a})
U_{0}(t_{a},0)   
=
\left(
\begin{array}{ccc}
0&  1& 0 \\
1&  0& 0 \\
0&  0& 1
\end{array}
\right).
\end{equation}

\vspace{5mm} 
Next we construct 
\[
\left(
\begin{array}{ccc}
0&  0& 1 \\
0&  1& 0 \\
1&  0& 0
\end{array}
\right).
\]
Similarly in the preceeding 
\begin{eqnarray}
&&U_{0}(t_{e},t_{d})U_{1}(t_{d},t_{c})U_{0}(t_{c},t_{b})U_{3}(t_{b},t_{a})
U_{0}(t_{a},0)         \nonumber \\
=&&
\mbox{e}^{-iE_{0}t_{e}}
\left(
  \begin{array}{ccc}
  1 &   &                                                                \\
    & \mbox{e}^{-i\{\omega_{1}(t_{d}-t_{c})+\Delta_{1}(t_{e}-t_{d}+t_{c})
    +\phi_{1}\}} &                                                       \\
    &   & \mbox{e}^{-i\{\omega_{3}(t_{b}-t_{a})+
    \Delta_{2}(t_{e}-t_{c})+\phi_{3}\}}
  \end{array}
\right)
\left(
\begin{array}{ccc}
0&  0& 1 \\
0&  1& 0 \\
1&  0& 0
\end{array}
\right)          \nonumber 
\end{eqnarray}
with
\[
\Delta_{2}t_{a}=(3/2)\pi,\ \  g(t_{b}-t_{a})=\pi/2,\ \ 
\Delta_{2}(t_{c}-t_{b})=(3/2)\pi,\ \ g(t_{d}-t_{c})=2\pi. 
\]

If we choose $t_{e}\ (> t_{d})$ as 
\[
E_{0}t_{e}=2\pi k \quad \mbox{for some}\ k\ \in \futon
\]
and the phases $\phi_{1},\ \phi_{3}$ as 
\[
\mbox{e}^{-i\{\omega_{1}(t_{d}-t_{c})+\Delta_{1}(t_{e}-t_{d}+t_{c})+\phi_{1}\}}
=1, \quad 
\mbox{e}^{-i\{\omega_{3}(t_{b}-t_{a})+\Delta_{2}(t_{e}-t_{c})+\phi_{3}\}}
=1, 
\]
then we finally obtain 
\begin{equation}
U_{0}(t_{e},t_{d})U_{1}(t_{d},t_{c})U_{0}(t_{c},t_{b})U_{3}(t_{b},t_{a})
U_{0}(t_{a},0)      
=
\left(
\begin{array}{ccc}
0&  0& 1 \\
0&  1& 0 \\
1&  0& 0
\end{array}
\right).         
\end{equation}

We note here that the matrix K in (\ref{eq:matrix-k}) can be constructed as 
\begin{equation}
\left(
\begin{array}{ccc}
0&  0& 1 \\
0&  1& 0 \\
1&  0& 0
\end{array}
\right)
\left(
\begin{array}{ccc}
0&  1& 0 \\
1&  0& 0 \\
0&  0& 1
\end{array}
\right)
\left(
\begin{array}{ccc}
0&  0& 1 \\
0&  1& 0 \\
1&  0& 0
\end{array}
\right)
=
\left(
\begin{array}{ccc}
1&  0& 0 \\
0&  0& 1 \\
0&  1& 0
\end{array}
\right)=K. 
\end{equation}

\vspace{5mm}
Next we construct 
\begin{equation}
\Sigma_{3}=
\left(
\begin{array}{ccc}
 1&       &             \\
  & \sigma&             \\
  &       & {\sigma}^2
\end{array}
\right)
\end{equation}
where $\sigma=\mbox{exp}(2\pi i/3)$. 

From (\ref{eq:unitary-1}) and (\ref{eq:unitary-3}) we choose 
$t_{a}$ as $gt_{a}=2\pi$ and $t_{b}\ (> t_{a})$ as $g(t_{b}-t_{a})=2\pi$ 
such as 
\[
U_{1}(t_{a},0)
=\mbox{e}^{-iE_{0}t_{a}}
\left(
  \begin{array}{ccc}
  1 & &                                           \\
    & \mbox{e}^{-i(\omega_{1}t_{a}+\phi_{1})} &   \\
    & & \mbox{e}^{-i\Delta_{2}t_{a}}
  \end{array}
\right)
\]
and 
\[
U_{3}(t_{b},t_{a})
=
\mbox{e}^{-iE_{0}(t_{b}-t_{a})}
\left(
  \begin{array}{ccc}
  1 &   &                                                   \\
    & \mbox{e}^{-i\Delta_{1}(t_{b}-t_{a})} &                \\
    &   & \mbox{e}^{-i\{\omega_{3}(t_{b}-t_{a})+\phi_{3}\}}
  \end{array}
\right), 
\]
we have 
\[
U_{3}(t_{b},t_{a})U_{1}(t_{a},0)
=
\mbox{e}^{-iE_{0}t_{b}}
\left(
  \begin{array}{ccc}
  1 &  &                                                                  \\
    & \mbox{e}^{-i\{\Delta_{1}(t_{b}-t_{a})+\omega_{1}t_{a}+\phi_{1}\}} & \\
    &  & \mbox{e}^{-i\{\omega_{3}(t_{b}-t_{a})+\Delta_{2}t_{a}+\phi_{3}\}}
  \end{array}
\right).
\]
As we cannot remove the phase $\mbox{e}^{-iE_{0}t_{b}}$ we multiply 
the above matrix by $U_{0}(t_{c},t_{b})$ in (\ref{eq:unitary-0}) 
\[
U_{0}(t_{c},t_{b})U_{3}(t_{b},t_{a})U_{1}(t_{a},0)
=
\mbox{e}^{-iE_{0}t_{c}}
\left(
  \begin{array}{ccc}
  1 &  &                                                                  \\
    & \mbox{e}^{-i\{\Delta_{1}(t_{c}-t_{a})+\omega_{1}t_{a}+\phi_{1}\}} & \\
    &  & \mbox{e}^{-i\{\omega_{3}(t_{b}-t_{a})+\Delta_{2}(t_{c}-t_{b}+t_{a})
    +\phi_{3}\}}
  \end{array}
\right).
\]
Here if we choose $t_{c}$ as 
\[
E_{0}t_{c}=2\pi k \quad \mbox{for some}\ k\ \in \futon
\]
and the phases $\phi_{1},\ \phi_{3}$ as 
\[
\mbox{e}^{-i\{\Delta_{1}(t_{c}-t_{a})+\omega_{1}t_{a}+\phi_{1}\}}= 
\mbox{e}^{\frac{2\pi i}{3}}=\sigma, \quad 
\mbox{e}^{-i\{\omega_{3}(t_{b}-t_{a})+\Delta_{2}(t_{c}-t_{b}+t_{a})+
\phi_{3}\}}=\mbox{e}^{\frac{4\pi i}{3}}=\sigma^{2},
\]
then we finally obtain 
\begin{equation}
U_{0}(t_{c},t_{b})U_{3}(t_{b},t_{a})U_{1}(t_{a},0)
=
\left(
\begin{array}{ccc}
 1&       &             \\
  & \sigma&             \\
  &       & {\sigma}^2
\end{array}
\right)
=\Sigma_{3}.
\end{equation}
Moreover if we choose the phases $\phi_{1},\ \phi_{3}$ as 
\[
\mbox{e}^{-i\{\Delta_{1}(t_{c}-t_{a})+\omega_{1}t_{a}+\phi_{1}\}}= i, \quad 
\mbox{e}^{-i\{\omega_{3}(t_{b}-t_{a})+\Delta_{2}(t_{c}-t_{b}+t_{a})+
\phi_{3}\}}=i,
\]
then we obtain 
\begin{equation}
\label{eq:useful-matrix}
I\equiv U_{0}(t_{c},t_{b})U_{3}(t_{b},t_{a})U_{1}(t_{a},0)
=
\left(
\begin{array}{ccc}
 1&   &     \\
  & i &     \\
  &   & i
\end{array}
\right)
\end{equation}
, which will become useful in the following. 

\vspace{5mm}
Lastly we construct the Walsh--Hadamard matrix 
\begin{equation}
W=\frac{1}{\sqrt{3}}
\left(
\begin{array}{ccc}
1&          1&     1      \\
1& \sigma^{2}& \sigma     \\
1& \sigma    & \sigma^{2}
\end{array}
\right)
=\frac{1}{\sqrt{3}}
\left(
\begin{array}{ccc}
1&          1&     1                               \\
1& \frac{-1-i\sqrt{3}}{2}& \frac{-1+i\sqrt{3}}{2}  \\
1& \frac{-1+i\sqrt{3}}{2}& \frac{-1-i\sqrt{3}}{2}
\end{array}
\right)
\end{equation}
because 
\[
\sigma=\mbox{e}^{\frac{2\pi i}{3}}=\frac{-1+i\sqrt{3}}{2}, \quad 
\sigma^{2}=\mbox{e}^{\frac{4\pi i}{3}}=\frac{-1-i\sqrt{3}}{2}. 
\]
This construction is not easy. As preliminaries let us construct the matrix 
\begin{equation}
\label{eq:matrix-I}
F=
\left(
\begin{array}{cc}
1&     \\
 &   \mbox{e}^{i\frac{\pi}{4}}
     \left(
       \begin{array}{cc} 
             \mbox{cos}(\frac{\pi}{4}) & -i\mbox{sin}(\frac{\pi}{4}) \\
             -i\mbox{sin}(\frac{\pi}{4}) & \mbox{cos}(\frac{\pi}{4})
       \end{array}
     \right)
\end{array}
\right).
\end{equation}

From (\ref{eq:unitary-2}) we choose $t_{a}$ as $gt_{a}=\pi/4$ such as 
\[
U_{2}(t_{a},0)
=\mbox{e}^{-iE_{0}t_{a}}
\left(
  \begin{array}{ccc}
   1 &   &                                                         \\
     & \mbox{e}^{-i\Delta_{1}t_{a}} &                              \\
     &   & \mbox{e}^{-i\{(\omega_{2}+\Delta_{1})t_{a}+\phi_{2}\}} 
  \end{array}
\right)
\left(
  \begin{array}{ccc}
  1 &  &                                                        \\
    & \mbox{cos}(\frac{\pi}{4}) & -i\mbox{sin}(\frac{\pi}{4})   \\
    & -i\mbox{sin}(\frac{\pi}{4}) & \mbox{cos}(\frac{\pi}{4})   
  \end{array}
\right). 
\]
Multiplying the above matrix by the matrix in (\ref{eq:unitary-1}) 
\[
U_{1}(t_{b},t_{a})
=\mbox{e}^{-iE_{0}(t_{b}-t_{a})}
\left(
  \begin{array}{ccc}
  1 & &                                                     \\
    & \mbox{e}^{-i\{\omega_{1}(t_{b}-t_{a})+\phi_{1}\}} &   \\
    & & \mbox{e}^{-i\Delta_{2}(t_{b}-t_{a})}
  \end{array}
\right)
\]
with $g(t_{b}-t_{a})=2\pi$ and by the matrix $U_{0}(t_{c},t_{b})$ in 
(\ref{eq:unitary-0}), we have  
\begin{eqnarray}
&&U_{0}(t_{c},t_{b})U_{1}(t_{b},t_{a})U_{2}(t_{a},0)   
=\mbox{e}^{-iE_{0}t_{c}}\times          \nonumber \\
&&\left(
  \begin{array}{ccc}
  1 & &                                                                    \\
    & \mbox{e}^{-i\{\Delta_{1}(t_{c}-t_{b}+t_{a})+
    \omega_{1}(t_{b}-t_{a})+\phi_{1}\}} &                                  \\
    & & \mbox{e}^{-i\{\Delta_{2}(t_{c}-t_{a})+(\omega_{2}+\Delta_{1})t_{a}+
    \phi_{2}\}}
  \end{array}
\right)
\left(
  \begin{array}{ccc}
  1 &  &                                                        \\
    & \mbox{cos}(\frac{\pi}{4}) & -i\mbox{sin}(\frac{\pi}{4})   \\
    & -i\mbox{sin}(\frac{\pi}{4}) & \mbox{cos}(\frac{\pi}{4})   
  \end{array}
\right).          \nonumber 
\end{eqnarray}
Here if we choose $t_{c}$ as 
\[
E_{0}t_{c}=2\pi k \quad \mbox{for some}\ k\ \in \futon
\]
and the phases $\phi_{1},\ \phi_{2}$ as 
\[
\mbox{e}^{-i\{\Delta_{1}(t_{c}-t_{b}+t_{a})+\omega_{1}(t_{b}-t_{a})+
\phi_{1}\}}=\mbox{e}^{i\frac{\pi}{4}}, \quad 
\mbox{e}^{-i\{(\omega_{2}+\Delta_{1})t_{a}+\Delta_{2}(t_{c}-t_{a})+
\phi_{2} \}}=\mbox{e}^{i\frac{\pi}{4}}
\]
then we finally obtain the desired matrix $F$. 

From (\ref{eq:unitary-5}) 
\begin{eqnarray}
U_{5}(t_{a},0)&=&\mbox{e}^{-iE_{0}t_{a}}
\left(
  \begin{array}{ccc}
  1 & &                                        \\
    & \mbox{e}^{-i(\omega_{1}t_{a}+\phi_{1})} &    \\
    & & \mbox{e}^{-i(\omega_{3}t_{a}+\phi_{3})}
  \end{array}
\right)\times        \nonumber \\
&&
\left(
  \begin{array}{ccc}
     \mbox{cos}(\sqrt{2}gt_{a})& 
    -i\frac{\mbox{sin}(\sqrt{2}gt_{a})}{\sqrt{2}}& 
    -i\frac{\mbox{sin}(\sqrt{2}gt_{a})}{\sqrt{2}}       \\
    -i\frac{\mbox{sin}(\sqrt{2}gt_{a})}{\sqrt{2}}& 
     \frac{1+\mbox{cos}(\sqrt{2}gt_{a})}{2}& 
     \frac{-1+\mbox{cos}(\sqrt{2}gt_{a})}{2}            \\
    -i\frac{\mbox{sin}(\sqrt{2}gt_{a})}{\sqrt{2}}& 
     \frac{-1+\mbox{cos}(\sqrt{2}gt_{a})}{2}& 
     \frac{1+\mbox{cos}(\sqrt{2}gt_{a})}{2}       
  \end{array}
\right)            \nonumber 
\end{eqnarray}
we choose $t_{a}$ as 
\[
\mbox{cos}(\sqrt{2}gt_{a})=\frac{1}{\sqrt{3}}\quad \mbox{and}\quad 
\mbox{sin}(\sqrt{2}gt_{a})=\frac{\sqrt{2}}{\sqrt{3}} 
\Longleftrightarrow 
gt_{a}=\frac{\theta}{\sqrt{2}}\quad \mbox{for some}\ \theta, 
\]
then 
\[
U_{5}(t_{a},0)=\mbox{e}^{-iE_{0}t_{a}}
\left(
  \begin{array}{ccc}
  1 & &                                            \\
    & \mbox{e}^{-i(\omega_{1}t_{a}+\phi_{1})} &    \\
    & & \mbox{e}^{-i(\omega_{3}t_{a}+\phi_{3})}
  \end{array}
\right)
\frac{1}{\sqrt{3}}
\left(
  \begin{array}{ccc}
     1& -i& -i                                       \\
    -i& \frac{1+\sqrt{3}}{2}& \frac{1-\sqrt{3}}{2}   \\
    -i& \frac{1-\sqrt{3}}{2}& \frac{1+\sqrt{3}}{2} 
  \end{array}
\right).           
\]
To remove the phase we multiply the above matrix by $U_{0}(t_{b},t_{a})$ 
in (\ref{eq:unitary-0}) 
\begin{eqnarray}
&&U_{0}(t_{b},t_{a})U_{5}(t_{a},0)      \nonumber \\
=&&\mbox{e}^{-iE_{0}t_{b}}
\left(
  \begin{array}{ccc}
  1 & &                                                                   \\
    & \mbox{e}^{-i\{\Delta_{1}(t_{b}-t_{a})+\omega_{1}t_{a}+\phi_{1}\}} & \\
    & & \mbox{e}^{-i\{\Delta_{2}(t_{b}-t_{a})+\omega_{3}t_{a}+\phi_{3}\}}
  \end{array}
\right)
\frac{1}{\sqrt{3}}
\left(
  \begin{array}{ccc}
     1& -i& -i                                       \\
    -i& \frac{1+\sqrt{3}}{2}& \frac{1-\sqrt{3}}{2}   \\
    -i& \frac{1-\sqrt{3}}{2}& \frac{1+\sqrt{3}}{2} 
  \end{array}
\right)                     \nonumber
\end{eqnarray}
, and choose $t_{b}\ (> t_{a})$ as 
\[
E_{0}t_{b}=2\pi k \quad \mbox{for some}\ k\ \in \futon
\]
and the phases $\phi_{1},\ \phi_{3}$ as 
\[
\mbox{e}^{-i\{\Delta_{1}(t_{b}-t_{a})+\omega_{1}t_{a}+\phi_{1}\}}=1, \quad 
\mbox{e}^{-i\{\Delta_{2}(t_{b}-t_{a})+\omega_{3}t_{a}+\phi_{3}\}}=1
\]
then we have 
\[
U_{0}(t_{b},t_{a})U_{5}(t_{a},0)=
\frac{1}{\sqrt{3}}
\left(
  \begin{array}{ccc}
     1& -i& -i                                       \\
    -i& \frac{1+\sqrt{3}}{2}& \frac{1-\sqrt{3}}{2}   \\
    -i& \frac{1-\sqrt{3}}{2}& \frac{1+\sqrt{3}}{2} 
  \end{array}
\right).
\]
Next we multiply the above matrix by the matrix I in (\ref{eq:useful-matrix}) 
to become 
\begin{equation}
\label{eq:matrix-II}
IU_{0}(t_{b},t_{a})U_{5}(t_{a},0)I=
\frac{1}{\sqrt{3}}
\left(
  \begin{array}{ccc}
     1 & 1 & 1                                          \\
     1 & \frac{-1-\sqrt{3}}{2}& \frac{-1+\sqrt{3}}{2}   \\
     1 & \frac{-1+\sqrt{3}}{2}& \frac{-1-\sqrt{3}}{2} 
  \end{array}
\right).
\end{equation}

For the unitary matrix 
\[
\mbox{e}^{\frac{\pi i}{4}}
\left(
\begin{array}{cc}
  \mbox{cos}(\frac{\pi}{4})& -i\mbox{sin}(\frac{\pi}{4}) \\
-i\mbox{sin}(\frac{\pi}{4})&   \mbox{cos}(\frac{\pi}{4})  
\end{array}
\right)
=
\frac{1+i}{\sqrt{2}}
\left(
\begin{array}{cc}
\frac{1}{\sqrt{2}} & \frac{-i}{\sqrt{2}} \\
\frac{-i}{\sqrt{2}}& \frac{1}{\sqrt{2}}  
\end{array}
\right)
=
\frac{1+i}{2}
\left(
\begin{array}{cc}
1& -i \\
-i& 1  
\end{array}
\right), 
\]
it is easy to check 
\begin{eqnarray}
&&\frac{1+i}{2}
\left(
\begin{array}{cc}
1& -i \\
-i& 1  
\end{array}
\right)
\left(
\begin{array}{cc}
\frac{-1-\sqrt{3}}{2}& \frac{-1+\sqrt{3}}{2} \\
\frac{-1+\sqrt{3}}{2}& \frac{-1-\sqrt{3}}{2} 
\end{array}
\right)
=
\left(
\begin{array}{cc}
\frac{-1-i\sqrt{3}}{2}& \frac{-1+i\sqrt{3}}{2}  \\
\frac{-1+i\sqrt{3}}{2}& \frac{-1-i\sqrt{3}}{2}
\end{array}
\right),       \nonumber \\
&&\frac{1+i}{2}
\left(
\begin{array}{cc}
1& -i \\
-i& 1  
\end{array}
\right)
\left(
\begin{array}{c}
1 \\
1  
\end{array}
\right)
=
\left(
\begin{array}{c}
1 \\
1  
\end{array}
\right).      \nonumber 
\end{eqnarray}

Therefore multiplying (\ref{eq:matrix-II}) by (\ref{eq:matrix-I}) 
we finally obtain 
\begin{eqnarray}
&&\left(
\begin{array}{cc}
1&     \\
 &   \mbox{e}^{i\frac{\pi}{4}}
     \left(
       \begin{array}{cc} 
             \mbox{cos}(\frac{\pi}{4}) & -i\mbox{sin}(\frac{\pi}{4})  \\
             -i\mbox{sin}(\frac{\pi}{4}) & \mbox{cos}(\frac{\pi}{4})
       \end{array}
     \right)
\end{array}
\right)
\frac{1}{\sqrt{3}}
\left(
  \begin{array}{ccc}
     1 & 1 & 1                                          \\
     1 & \frac{-1-\sqrt{3}}{2}& \frac{-1+\sqrt{3}}{2}   \\
     1 & \frac{-1+\sqrt{3}}{2}& \frac{-1-\sqrt{3}}{2} 
  \end{array}
\right)        \nonumber   \\
=
&&\frac{1}{\sqrt{3}}
\left(
\begin{array}{ccc}
1&          1&     1                               \\
1& \frac{-1-i\sqrt{3}}{2}& \frac{-1+i\sqrt{3}}{2}  \\
1& \frac{-1+i\sqrt{3}}{2}& \frac{-1-i\sqrt{3}}{2}
\end{array}
\right)
=
\frac{1}{\sqrt{3}}
\left(
\begin{array}{ccc}
1&          1&     1      \\
1& \sigma^{2}& \sigma     \\
1& \sigma    & \sigma^{2}
\end{array}
\right)
=W.
\end{eqnarray}

As shown in the proof, to construct the Walsh--Hadamard matrix $W$ in terms of 
Rabi oscillations is not easy.

\subsection{Quantum Logic Gates on N Level Systems $\cdots$ Problems}

As in the three level case it is easy to construct generalized Pauli matrices 
$\Sigma_{1}, \Sigma_{3}$ (\ref{eq:gener-pauli}). For example, 
in the four level one we can construct $\Sigma_{1},\ \Sigma_{3}$ like 
\[
\Sigma_{1}=
\left(
\begin{array}{cccc}
 0 &   &   & 1 \\
   & 1 &   &   \\
   &   & 1 &   \\
 1 &   &   & 0
\end{array}
\right)
\left(
\begin{array}{cccc}
 0 &   & 1 &   \\
   & 1 &   &   \\
 1 &   & 0 &   \\
   &   &   & 1
\end{array}
\right)
\left(
\begin{array}{cccc}
 0 & 1 &   &   \\
 1 & 0 &   &   \\
   &   & 1 &   \\
   &   &   & 1
\end{array}
\right)
\]
and 
\[
\Sigma_{3}=
\left(
\begin{array}{cccc}
 1 &   &   &            \\
   & 1 &   &            \\
   &   & 1 &            \\
   &   &   & \sigma^{3}
\end{array}
\right)
\left(
\begin{array}{cccc}
 1 &   &            &    \\
   & 1 &            &    \\
   &   & \sigma^{2} &    \\
   &   &            & 1
\end{array}
\right)
\left(
\begin{array}{cccc}
 1 &        &   &    \\
   & \sigma &   &    \\
   &        & 1 &    \\
   &        &   & 1
\end{array}
\right).
\]

However, it is not easy to construct the generalized Walsh--Hadamard matrix 
$W$ (\ref{eq:Large-double}). In fact, we don't know how to construct it, 
so we present 

\begin{flushleft}
{\bf Problem} \quad Construct the generalized Walsh--Hadamard matrix 
by making use of Rabi oscillations (in a general level system). 
\end{flushleft}
or, as an easier version, 

\begin{flushleft}
{\bf Problem} \quad Construct the generalized Walsh--Hadamard matrix 
by making use of Rabi oscillations in four level systems. 
\end{flushleft}
That is, 
\begin{equation}
W=\frac{1}{2}
\left(
\begin{array}{cccc}
 1 &  1 &  1 &  1   \\
 1 & -i & -1 &  i   \\
 1 & -1 &  1 & -1   \\
 1 &  i & -1 & -i
\end{array}
\right)
\end{equation}
since $\sigma=\mbox{exp}(\frac{\pi i}{2})=i$.

\subsection{Formal Construction of U(3)}

Here we give a formal construction to $SU(3)$ by using a generalization of 
the Euler angle parametrization in $SU(2)$ by \cite{MB}, \cite{TSu} and 
next give a construction to $U(3)$ by adding phases. 

It is known that any element $U$ in $SU(3)$ can be written as 
\begin{equation}
U\equiv U(\alpha,\beta,\gamma,\theta,a,b,c,\phi)=
\mbox{e}^{i\alpha \lambda_{3}}
\mbox{e}^{i\beta  \lambda_{2}}
\mbox{e}^{i\gamma \lambda_{3}}
\mbox{e}^{i\theta \lambda_{5}}
\mbox{e}^{ia     \lambda_{3}}
\mbox{e}^{ib     \lambda_{2}}
\mbox{e}^{ic     \lambda_{3}}
\mbox{e}^{i\phi   \lambda_{8}}
\end{equation}
by using a generalization of Euler angles $(\alpha, \beta, \gamma, \theta, 
a, b, c, \phi)$, where $\lambda_{2}, \lambda_{3}, \lambda_{5}, \lambda_{8}$ 
are well--known Gell-Mann matrices defined by 
\begin{eqnarray}
&&
\lambda_{2}=
\left(
\begin{array}{ccc}
 0 & -i & 0   \\
 i &  0 & 0   \\
 0 &  0 & 0   \\
\end{array}
\right),\
\lambda_{3}=
\left(
\begin{array}{ccc}
 1 & 0  & 0   \\
 0 & -1 & 0   \\
 0 &  0 & 0   \\
\end{array}
\right),\
\lambda_{5}=
\left(
\begin{array}{ccc}
 0 & 0 & -i   \\
 0 & 0 &  0   \\
 i & 0 &  0   \\
\end{array}
\right),\
\lambda_{8}=\frac{1}{\sqrt{3}}
\left(
\begin{array}{ccc}
 1 & 0 &  0   \\
 0 & 1 &  0   \\
 0 & 0 & -2   \\
\end{array}
\right).    \nonumber \\
&&{}
\end{eqnarray}
Explicitly written 
\begin{eqnarray}
&&\mbox{e}^{i\alpha \lambda_{3}}
=
\left(
\begin{array}{ccc}
 \mbox{e}^{i\alpha} & 0  & 0   \\
 0 & \mbox{e}^{-i\alpha} & 0   \\
        0 &            0 & 1   \\
\end{array}
\right),\qquad 
\mbox{e}^{i\beta  \lambda_{2}}
=
\left(
\begin{array}{ccc}
  \mbox{cos}(\beta) & \mbox{sin}(\beta) & 0   \\
 -\mbox{sin}(\beta) & \mbox{cos}(\beta) & 0   \\
        0           &  0                & 1   \\
\end{array}
\right),         \nonumber \\
&&\mbox{e}^{i\theta \lambda_{5}}
=
\left(
\begin{array}{ccc}
  \mbox{cos}(\theta) & 0 & \mbox{sin}(\theta)   \\
                   0 & 1 &  0                   \\
 -\mbox{sin}(\theta) & 0 & \mbox{cos}(\theta)  \\
\end{array}
\right),\qquad 
\mbox{e}^{i\phi   \lambda_{8}}
=
\left(
\begin{array}{ccc}
 \mbox{e}^{i\phi/\sqrt{3}} & 0 &  0    \\
 0 & \mbox{e}^{i\phi/\sqrt{3}} &  0    \\
 0 & 0 & \mbox{e}^{-2i\phi/\sqrt{3}}   \\
\end{array}
\right).   
\end{eqnarray}

It is not difficult to construct the above matrices by making use of 
unitary operations $U_{0}(t,0)$ $\sim$ $U_{7}(t,0)$ in section 3. 
We leave it to the readers. 

\begin{flushleft}
{\bf Exercise}\quad Construct them. 
\end{flushleft}

Any element $U$ in $U(3)$ can be obtained by multiplying an element in 
$SU(3)$ by a phase matrix, for example 
\[
\left(
\begin{array}{ccc}
 1 & 0 & 0                     \\
 0 & \mbox{e}^{i\epsilon} & 0  \\
 0 & 0 & \mbox{e}^{i\delta}
\end{array}
\right). 
\]
Then it is easy to construct it, so we finish a formal construction of 
$U(3)$. 

{\bf A comment is in order}.\ This construction is only formal, so it is 
not useful in a realistic scene. For example, when we want to construct the 
Walsh--Hadamard matrix with this form, it is almost impossible to find 
the angles above.

\section{Discussion}

We have given the explicit constructions to the generalized Pauli matrices 
and generalized Walsh--Hadamard matrix in the three level systems 
by making use of Rabi oscillations. Therefore we can perform important 
quantum logic gates used in \cite{CBKG}, \cite{BGS}, \cite{KBB}, \cite{KF4} 
at least in the three level systems. 

For a qudit theory it is enough for us to consider three and four level 
systems. However we have not suceeded in constructing the generalized 
Walsh--Hadamard matrix in four level systems. We leave it to the readers. 

We conclude this paper by expecting that some experimentalists in quantum 
optics will check our method.

\newpage 

\begin{center}
\begin{Large}
\noindent{\bfseries Appendix}
\end{Large}
\end{center}

\vspace{5mm}
In this appendix we give slightly generalized versions of the contents in 
subsections 3.5, 3.6, 3.7, 3.8.  That is, we consider models with two 
coupling constants $g_{1},\ g_{2}$ and solve them exactly. 
Unfortunately in the case of VII (three coupling constants) we cannot give 
an explicit solution owing to some technical difficulty.

\begin{flushleft}
\begin{Large}
{\bf Unitary Transformation of type {IV}}
\end{Large}
\end{flushleft}

The Hamiltonian that we are treating is 
\begin{eqnarray}
\label{eq:hamiltonian-4-g}
H_{IV}&=&
\left(
\begin{array}{ccc}
   E_{0} & g_{1}\mbox{e}^{i(\phi_{1}+\omega_{1}t)} & 0     \\
   g_{1}\mbox{e}^{-i(\phi_{1}+\omega_{1}t)} & E_{1} & 
   g_{2}\mbox{e}^{i(\phi_{2}+\omega_{2}t)}                 \\
    0 & g_{2}\mbox{e}^{-i(\phi_{2}+\omega_{2}t)} & E_{2}
  \end{array}
\right)             \nonumber \\
&=&
E_{0}{\bf 1}_{3}+
\left(
  \begin{array}{ccc}
   0 & g_{1}\mbox{e}^{i(\phi_{1}+\omega_{1}t)} & 0            \\
   g_{1}\mbox{e}^{-i(\phi_{1}+\omega_{1}t)} & \Delta_{1} & 
   g_{2}\mbox{e}^{i(\phi_{2}+\omega_{2}t)}                    \\
    0 & g_{2}\mbox{e}^{-i(\phi_{2}+\omega_{2}t)} & \Delta_{2}
  \end{array}
\right).
\end{eqnarray}

We would like to solve the Schr{\" o}dinger equation 
\begin{equation}
\label{eq:schrodinger-equation-4-g}
i\frac{d}{dt}\Psi=H_{IV}\Psi.
\end{equation}

If we note the following decomposition 
\begin{eqnarray}
&&\left(
  \begin{array}{ccc}
   0 & g_{1}\mbox{e}^{i(\phi_{1}+\omega_{1}t)} & 0           \\
   g_{1}\mbox{e}^{-i(\phi_{1}+\omega_{1}t)} & \Delta_{1} & 
   g_{2}\mbox{e}^{i(\phi_{2}+\omega_{2}t)}                   \\
   0 & g_{2}\mbox{e}^{-i(\phi_{2}+\omega_{2}t)} & \Delta_{2}
  \end{array}
\right)
=               \nonumber \\
&&\left(
  \begin{array}{ccc}
  1 & &                                                           \\
    & \mbox{e}^{-i(\phi_{1}+\omega_{1}t)} &                       \\
    & & \mbox{e}^{-i(\phi_{1}+\omega_{1}t+\phi_{2}+\omega_{2}t)}
  \end{array}
\right)
\left(
  \begin{array}{ccc}
   0 & g_{1} & 0                       \\
   g_{1} & \Delta_{1} & g_{2}   \\
   0 & g_{2} & \Delta_{2}
  \end{array}
\right)
\left(
  \begin{array}{ccc}
  1 & &                                                          \\
    & \mbox{e}^{i(\phi_{1}+\omega_{1}t)} &                       \\
    & & \mbox{e}^{i(\phi_{1}+\omega_{1}t+\phi_{2}+\omega_{2}t)}
  \end{array}
\right)         \nonumber \\ 
&&{}  
\end{eqnarray}
and define 
\begin{equation}
\Phi=\mbox{e}^{itE_{0}}
\left(
  \begin{array}{ccc}
  1 & &                                                          \\
    & \mbox{e}^{i(\phi_{1}+\omega_{1}t)} &                       \\
    & & \mbox{e}^{i(\phi_{1}+\omega_{1}t+\phi_{2}+\omega_{2}t)}
  \end{array}
\right)\Psi 
\end{equation}
, then it is not difficult to see 
\begin{equation}
i\frac{d}{dt}\Phi=
\left(
  \begin{array}{ccc}
   0 & g_{1} & 0                                  \\
   g_{1} & \Delta_{1}-\omega_{1} & g_{2}          \\
   0 & g_{2} & \Delta_{2}-\omega_{1}-\omega_{2}
  \end{array}
\right)\Phi
\equiv 
\tilde{H}_{IV}\Phi.
\end{equation}
Here we set the resonance condition 
\begin{equation}
\label{eq:resonance-4-g}
  \Delta_{1}=\omega_{1},\quad \Delta_{2}=\omega_{1}+\omega_{2}, 
\end{equation}
so the solution is obtained to be 
\begin{eqnarray}
\Phi(t)&=&\mbox{exp}(-it\tilde{H}_{IV})\Phi(0)=     \nonumber \\
&=&
\left(
  \begin{array}{ccc}
  \frac{g_{1}^{2}\mbox{cos}(\sqrt{g_{1}^{2}+g_{2}^{2}}t)+g_{2}^{2}}
  {g_{1}^{2}+g_{2}^{2}}& 
 -i\frac{g_{1}\mbox{sin}(\sqrt{g_{1}^{2}+g_{2}^{2}}t)}
  {\sqrt{g_{1}^{2}+g_{2}^{2}}}& 
  \frac{g_{1}g_{2}\mbox{cos}(\sqrt{g_{1}^{2}+g_{2}^{2}}t)-g_{1}g_{2}}
  {g_{1}^{2}+g_{2}^{2}}  \\
 -i\frac{g_{1}\mbox{sin}(\sqrt{g_{1}^{2}+g_{2}^{2}}t)}
  {\sqrt{g_{1}^{2}+g_{2}^{2}}}& \mbox{cos}(\sqrt{g_{1}^{2}+g_{2}^{2}}t)&
 -i\frac{g_{2}\mbox{sin}(\sqrt{g_{1}^{2}+g_{2}^{2}}t)}
  {\sqrt{g_{1}^{2}+g_{2}^{2}}} \\
  \frac{g_{1}g_{2}\mbox{cos}(\sqrt{g_{1}^{2}+g_{2}^{2}}t)-g_{1}g_{2}}
  {g_{1}^{2}+g_{2}^{2}}&
  -i\frac{g_{2}\mbox{sin}(\sqrt{g_{1}^{2}+g_{2}^{2}}t)}
  {\sqrt{g_{1}^{2}+g_{2}^{2}}}&
  \frac{g_{2}^{2}\mbox{cos}(\sqrt{g_{1}^{2}+g_{2}^{2}}t)+g_{1}^{2}}
  {g_{1}^{2}+g_{2}^{2}} 
  \end{array}
\right)\Phi(0).
\end{eqnarray}

As a result the solution we are looking for is 
\begin{eqnarray}
\Psi(t)&=&\mbox{e}^{-itE_{0}}
\left(
  \begin{array}{ccc}
  1 & &                                                           \\
    & \mbox{e}^{-i(\phi_{1}+\omega_{1}t)} &                       \\
    & & \mbox{e}^{-i(\phi_{1}+\omega_{1}t+\phi_{2}+\omega_{2}t)}
  \end{array}
\right)\times             \nonumber \\
&&\left(
  \begin{array}{ccc}
  \frac{g_{1}^{2}\mbox{cos}(\sqrt{g_{1}^{2}+g_{2}^{2}}t)+g_{2}^{2}}
  {g_{1}^{2}+g_{2}^{2}}& 
 -i\frac{g_{1}\mbox{sin}(\sqrt{g_{1}^{2}+g_{2}^{2}}t)}
  {\sqrt{g_{1}^{2}+g_{2}^{2}}}& 
  \frac{g_{1}g_{2}\mbox{cos}(\sqrt{g_{1}^{2}+g_{2}^{2}}t)-g_{1}g_{2}}
  {g_{1}^{2}+g_{2}^{2}}  \\
 -i\frac{g_{1}\mbox{sin}(\sqrt{g_{1}^{2}+g_{2}^{2}}t)}
  {\sqrt{g_{1}^{2}+g_{2}^{2}}}& \mbox{cos}(\sqrt{g_{1}^{2}+g_{2}^{2}}t)&
 -i\frac{g_{2}\mbox{sin}(\sqrt{g_{1}^{2}+g_{2}^{2}}t)}
  {\sqrt{g_{1}^{2}+g_{2}^{2}}} \\
  \frac{g_{1}g_{2}\mbox{cos}(\sqrt{g_{1}^{2}+g_{2}^{2}}t)-g_{1}g_{2}}
  {g_{1}^{2}+g_{2}^{2}}&
  -i\frac{g_{2}\mbox{sin}(\sqrt{g_{1}^{2}+g_{2}^{2}}t)}
  {\sqrt{g_{1}^{2}+g_{2}^{2}}}&
  \frac{g_{2}^{2}\mbox{cos}(\sqrt{g_{1}^{2}+g_{2}^{2}}t)+g_{1}^{2}}
  {g_{1}^{2}+g_{2}^{2}} 
  \end{array}
\right)\Phi(0).
\end{eqnarray}

For the latter use we set 
\begin{eqnarray}
\widetilde{U}_{4}(t,0)&=&\mbox{e}^{-itE_{0}}
\left(
  \begin{array}{ccc}
  1 & &                                                           \\
    & \mbox{e}^{-i(\phi_{1}+\omega_{1}t)} &                       \\
    & & \mbox{e}^{-i(\phi_{1}+\omega_{1}t+\phi_{2}+\omega_{2}t)}
  \end{array}
\right)\times             \nonumber \\
&&\left(
  \begin{array}{ccc}
  \frac{g_{1}^{2}\mbox{cos}(\sqrt{g_{1}^{2}+g_{2}^{2}}t)+g_{2}^{2}}
  {g_{1}^{2}+g_{2}^{2}}& 
 -i\frac{g_{1}\mbox{sin}(\sqrt{g_{1}^{2}+g_{2}^{2}}t)}
  {\sqrt{g_{1}^{2}+g_{2}^{2}}}& 
  \frac{g_{1}g_{2}\mbox{cos}(\sqrt{g_{1}^{2}+g_{2}^{2}}t)-g_{1}g_{2}}
  {g_{1}^{2}+g_{2}^{2}}  \\
 -i\frac{g_{1}\mbox{sin}(\sqrt{g_{1}^{2}+g_{2}^{2}}t)}
  {\sqrt{g_{1}^{2}+g_{2}^{2}}}& \mbox{cos}(\sqrt{g_{1}^{2}+g_{2}^{2}}t)&
 -i\frac{g_{2}\mbox{sin}(\sqrt{g_{1}^{2}+g_{2}^{2}}t)}
  {\sqrt{g_{1}^{2}+g_{2}^{2}}} \\
  \frac{g_{1}g_{2}\mbox{cos}(\sqrt{g_{1}^{2}+g_{2}^{2}}t)-g_{1}g_{2}}
  {g_{1}^{2}+g_{2}^{2}}&
  -i\frac{g_{2}\mbox{sin}(\sqrt{g_{1}^{2}+g_{2}^{2}}t)}
  {\sqrt{g_{1}^{2}+g_{2}^{2}}}&
  \frac{g_{2}^{2}\mbox{cos}(\sqrt{g_{1}^{2}+g_{2}^{2}}t)+g_{1}^{2}}
  {g_{1}^{2}+g_{2}^{2}} 
  \end{array}
\right).
\end{eqnarray}

\vspace{5mm}

\begin{flushleft}
\begin{Large}
{\bf Unitary Transformation of type {V}}
\end{Large}
\end{flushleft}

The Hamiltonian is 
\begin{eqnarray}
\label{eq:hamiltonian-5-g}
H_{V}&=&
\left(
  \begin{array}{ccc}
   E_{0} & g_{1}\mbox{e}^{i(\phi_{1}+\omega_{1}t)} & 
   g_{3}\mbox{e}^{i(\phi_{3}+\omega_{3}t)}                \\
   g_{1}\mbox{e}^{-i(\phi_{1}+\omega_{1}t)} & E_{1} & 0   \\
   g_{3}\mbox{e}^{-i(\phi_{3}+\omega_{3}t)}& 0 & E_{2}
  \end{array}
\right)            \nonumber \\
&=& E_{0}{\bf 1}_{3}+
\left(
  \begin{array}{ccc}
   0 & g_{1}\mbox{e}^{i(\phi_{1}+\omega_{1}t)} & 
   g_{3}\mbox{e}^{i(\phi_{3}+\omega_{3}t)}                     \\
   g_{1}\mbox{e}^{-i(\phi_{1}+\omega_{1}t)} & \Delta_{1} & 0   \\
   g_{3}\mbox{e}^{-i(\phi_{3}+\omega_{3}t)}& 0 & \Delta_{2}
  \end{array}
\right).
\end{eqnarray}

We would like to solve the Schr{\" o}dinger equation 
\begin{equation}
\label{eq:schrodinger-equation-5-g}
i\frac{d}{dt}\Psi=H_{V}\Psi.
\end{equation}

If we note the following decomposition 
\begin{eqnarray}
&&\left(
  \begin{array}{ccc}
   0 & g_{1}\mbox{e}^{i(\phi_{1}+\omega_{1}t)} & 
   g_{3}\mbox{e}^{i(\phi_{3}+\omega_{3}t)}                     \\
   g_{1}\mbox{e}^{-i(\phi_{1}+\omega_{1}t)} & \Delta_{1} & 0   \\
   g_{3}\mbox{e}^{-i(\phi_{3}+\omega_{3}t)}& 0 & \Delta_{2}
  \end{array}
\right)
=               \nonumber \\
&&\left(
  \begin{array}{ccc}
  1 &   &                                          \\
    & \mbox{e}^{-i(\phi_{1}+\omega_{1}t)}     &    \\
    &   & \mbox{e}^{-i(\phi_{3}+\omega_{3}t)}
  \end{array}
\right)
\left(
  \begin{array}{ccc}
   0 & g_{1} & g_{3}         \\
   g_{1} & \Delta_{1} & 0    \\
   g_{3} & 0 & \Delta_{2}
  \end{array}
\right)
\left(
  \begin{array}{ccc}
  1 &  &                                         \\
    & \mbox{e}^{i(\phi_{1}+\omega_{1}t)}   &     \\
    & & \mbox{e}^{i(\phi_{3}+\omega_{3}t)}
  \end{array}
\right)         \nonumber \\ 
&&{}  
\end{eqnarray}
and define 
\begin{equation}
\Phi=\mbox{e}^{itE_{0}}
\left(
  \begin{array}{ccc}
  1 & &                                        \\
    & \mbox{e}^{i(\phi_{1}+\omega_{1}t)} &     \\
    & & \mbox{e}^{i(\phi_{3}+\omega_{3}t)}
  \end{array}
\right)\Psi 
\end{equation}
, then it is not difficult to see 
\begin{equation}
i\frac{d}{dt}\Phi=
\left(
  \begin{array}{ccc}
   0 & g_{1} & g_{3}                    \\
   g_{1} & \Delta_{1}-\omega_{1} & 0    \\
   g_{3} & 0 & \Delta_{2}-\omega_{3}
  \end{array}
\right)\Phi
\equiv 
\tilde{H}_{V}\Phi.
\end{equation}
Here we set the resonance condition 
\begin{equation}
\label{eq:resonance-5-g}
  \Delta_{1}=\omega_{1},\quad \Delta_{2}=\omega_{3}, 
\end{equation}
so the solution is obtained to be 
\begin{eqnarray}
\Phi(t)&=&\mbox{exp}(-it\tilde{H}_{IV})\Phi(0)=          \nonumber \\
&=&
\left(
  \begin{array}{ccc}
  \mbox{cos}(\sqrt{g_{1}^{2}+g_{3}^{2}}t)&
 -i\frac{g_{1}\mbox{sin}(\sqrt{g_{1}^{2}+g_{3}^{2}}t)}
  {\sqrt{g_{1}^{2}+g_{3}^{2}}}& 
 -i\frac{g_{3}\mbox{sin}(\sqrt{g_{1}^{2}+g_{3}^{2}}t)}
  {\sqrt{g_{1}^{2}+g_{3}^{2}}}                                        \\
 -i\frac{g_{1}\mbox{sin}(\sqrt{g_{1}^{2}+g_{3}^{2}}t)}
  {\sqrt{g_{1}^{2}+g_{3}^{2}}}& 
  \frac{g_{1}^{2}\mbox{cos}(\sqrt{g_{1}^{2}+g_{3}^{2}}t)+g_{3}^{2}}
  {g_{1}^{2}+g_{3}^{2}}& 
  \frac{g_{1}g_{3}\mbox{cos}(\sqrt{g_{1}^{2}+g_{3}^{2}}t)-g_{1}g_{3}}
  {g_{1}^{2}+g_{3}^{2}}                                                \\
  -i\frac{g_{3}\mbox{sin}(\sqrt{g_{1}^{2}+g_{3}^{2}}t)}
  {\sqrt{g_{1}^{2}+g_{3}^{2}}}&       
  \frac{g_{1}g_{3}\mbox{cos}(\sqrt{g_{1}^{2}+g_{3}^{2}}t)-g_{1}g_{3}}
  {g_{1}^{2}+g_{3}^{2}}&
  \frac{g_{3}^{2}\mbox{cos}(\sqrt{g_{1}^{2}+g_{3}^{2}}t)+g_{1}^{2}}
  {g_{1}^{2}+g_{3}^{2}} 
  \end{array}
\right)\Phi(0).
\end{eqnarray}

As a result the solution we are looking for is 
\begin{eqnarray}
\Psi(t)&=&\mbox{e}^{-itE_{0}}
\left(
  \begin{array}{ccc}
  1 & &                                         \\
    & \mbox{e}^{-i(\phi_{1}+\omega_{1}t)} &     \\
    & & \mbox{e}^{-i(\phi_{3}+\omega_{3}t)}
  \end{array}
\right)\times         \nonumber \\
&&
\left(
  \begin{array}{ccc}
  \mbox{cos}(\sqrt{g_{1}^{2}+g_{3}^{2}}t)&
 -i\frac{g_{1}\mbox{sin}(\sqrt{g_{1}^{2}+g_{3}^{2}}t)}
  {\sqrt{g_{1}^{2}+g_{3}^{2}}}& 
 -i\frac{g_{3}\mbox{sin}(\sqrt{g_{1}^{2}+g_{3}^{2}}t)}
  {\sqrt{g_{1}^{2}+g_{3}^{2}}}                                        \\
 -i\frac{g_{1}\mbox{sin}(\sqrt{g_{1}^{2}+g_{3}^{2}}t)}
  {\sqrt{g_{1}^{2}+g_{3}^{2}}}& 
  \frac{g_{1}^{2}\mbox{cos}(\sqrt{g_{1}^{2}+g_{3}^{2}}t)+g_{3}^{2}}
  {g_{1}^{2}+g_{3}^{2}}& 
  \frac{g_{1}g_{3}\mbox{cos}(\sqrt{g_{1}^{2}+g_{3}^{2}}t)-g_{1}g_{3}}
  {g_{1}^{2}+g_{3}^{2}}                                                \\
  -i\frac{g_{3}\mbox{sin}(\sqrt{g_{1}^{2}+g_{3}^{2}}t)}
  {\sqrt{g_{1}^{2}+g_{3}^{2}}}&       
  \frac{g_{1}g_{3}\mbox{cos}(\sqrt{g_{1}^{2}+g_{3}^{2}}t)-g_{1}g_{3}}
  {g_{1}^{2}+g_{3}^{2}}&
  \frac{g_{3}^{2}\mbox{cos}(\sqrt{g_{1}^{2}+g_{3}^{2}}t)+g_{1}^{2}}
  {g_{1}^{2}+g_{3}^{2}} 
  \end{array}
\right)\Phi(0).
\end{eqnarray}

For the latter use we set 
\begin{eqnarray}
\widetilde{U}_{5}(t,0)&=&\mbox{e}^{-itE_{0}}
\left(
  \begin{array}{ccc}
  1 & &                                         \\
    & \mbox{e}^{-i(\phi_{1}+\omega_{1}t)} &     \\
    & & \mbox{e}^{-i(\phi_{3}+\omega_{3}t)}
  \end{array}
\right)\times         \nonumber \\
&&
\left(
  \begin{array}{ccc}
  \mbox{cos}(\sqrt{g_{1}^{2}+g_{3}^{2}}t)&
 -i\frac{g_{1}\mbox{sin}(\sqrt{g_{1}^{2}+g_{3}^{2}}t)}
  {\sqrt{g_{1}^{2}+g_{3}^{2}}}& 
 -i\frac{g_{3}\mbox{sin}(\sqrt{g_{1}^{2}+g_{3}^{2}}t)}
  {\sqrt{g_{1}^{2}+g_{3}^{2}}}                                        \\
 -i\frac{g_{1}\mbox{sin}(\sqrt{g_{1}^{2}+g_{3}^{2}}t)}
  {\sqrt{g_{1}^{2}+g_{3}^{2}}}& 
  \frac{g_{1}^{2}\mbox{cos}(\sqrt{g_{1}^{2}+g_{3}^{2}}t)+g_{3}^{2}}
  {g_{1}^{2}+g_{3}^{2}}& 
  \frac{g_{1}g_{3}\mbox{cos}(\sqrt{g_{1}^{2}+g_{3}^{2}}t)-g_{1}g_{3}}
  {g_{1}^{2}+g_{3}^{2}}                                                \\
  -i\frac{g_{3}\mbox{sin}(\sqrt{g_{1}^{2}+g_{3}^{2}}t)}
  {\sqrt{g_{1}^{2}+g_{3}^{2}}}&       
  \frac{g_{1}g_{3}\mbox{cos}(\sqrt{g_{1}^{2}+g_{3}^{2}}t)-g_{1}g_{3}}
  {g_{1}^{2}+g_{3}^{2}}&
  \frac{g_{3}^{2}\mbox{cos}(\sqrt{g_{1}^{2}+g_{3}^{2}}t)+g_{1}^{2}}
  {g_{1}^{2}+g_{3}^{2}} 
  \end{array}
\right).
\end{eqnarray}

\vspace{5mm}

\begin{flushleft}
\begin{Large}
{\bf Unitary Transformation of type {VI}}
\end{Large}
\end{flushleft}

The Hamiltonian is 
\begin{eqnarray}
\label{eq:hamiltonian-6-g}
H_{VI}&=&
\left(
  \begin{array}{ccc}
   E_{0} & 0 & g_{3}\mbox{e}^{i(\phi_{3}+\omega_{3}t)}   \\
   0 & E_{1} & g_{2}\mbox{e}^{i(\phi_{2}+\omega_{2}t)}   \\
   g_{3}\mbox{e}^{-i(\phi_{3}+\omega_{3}t)} & 
   g_{2}\mbox{e}^{-i(\phi_{2}+\omega_{2}t)} & E_{2}
  \end{array}
\right)             \nonumber \\
&=& E_{0}{\bf 1}_{3}+
\left(
  \begin{array}{ccc}
   0 & 0 & g_{3}\mbox{e}^{i(\phi_{3}+\omega_{3}t)}            \\
   0 & \Delta_{1} & g_{2}\mbox{e}^{i(\phi_{2}+\omega_{2}t)}   \\
   g_{3}\mbox{e}^{-i(\phi_{3}+\omega_{3}t)} & 
   g_{2}\mbox{e}^{-i(\phi_{2}+\omega_{2}t)} & \Delta_{2}
  \end{array}
\right).
\end{eqnarray}

We would like to solve the Schr{\" o}dinger equation 
\begin{equation}
\label{eq:schrodinger-equation-6-g}
i\frac{d}{dt}\Psi=H_{VI}\Psi.
\end{equation}

If we note the following decomposition 
\begin{eqnarray}
&&\left(
  \begin{array}{ccc}
   0 & 0 & g_{3}\mbox{e}^{i(\phi_{3}+\omega_{3}t)}            \\
   0 & \Delta_{1} & g_{2}\mbox{e}^{i(\phi_{2}+\omega_{2}t)}   \\
   g_{3}\mbox{e}^{-i(\phi_{3}+\omega_{3}t)} & 
   g_{2}\mbox{e}^{-i(\phi_{2}+\omega_{2}t)} & \Delta_{2}
  \end{array}
\right)
=               \nonumber \\
&&\left(
  \begin{array}{ccc} 
  1 & &                                                           \\
    & \mbox{e}^{-i(\phi_{3}+\omega_{3}t-\phi_{2}-\omega_{2}t)} &  \\
    & & \mbox{e}^{-i(\phi_{3}+\omega_{3}t)}
  \end{array}
\right)
\left(
  \begin{array}{ccc}
   0 & 0 & g_{3}                \\
   0 & \Delta_{1} & g_{2}       \\
   g_{3} & g_{2} & \Delta_{2}
  \end{array}
\right)
\left(
  \begin{array}{ccc}
  1 & &                                                          \\
    & \mbox{e}^{i(\phi_{3}+\omega_{3}t-\phi_{2}-\omega_{2}t)} &  \\
    & & \mbox{e}^{i(\phi_{3}+\omega_{3}t)} 
  \end{array}
\right)         \nonumber \\ 
&&{}  
\end{eqnarray}
and define 
\begin{equation}
\Phi=
\left(
  \begin{array}{ccc}
  1 & &                                                          \\
    & \mbox{e}^{i(\phi_{3}+\omega_{3}t-\phi_{2}-\omega_{2}t)} &  \\
    & & \mbox{e}^{i(\phi_{3}+\omega_{3}t)}
  \end{array}
\right)\Psi 
\end{equation}
, then it is not difficult to see 
\begin{equation}
i\frac{d}{dt}\Phi=
\left(
  \begin{array}{ccc}
   0 & 0 & g_{3}                                  \\
   0 & \Delta_{1}-\omega_{3}+ \omega_{2} & g_{2}  \\
   g_{3} & g_{2} & \Delta_{2}-\omega_{3}
  \end{array}
\right)\Phi
\equiv 
\tilde{H}_{VI}\Phi.
\end{equation}
Here we set the resonance condition 
\begin{equation}
\label{eq:resonance-6-g}
  \Delta_{1}=\omega_{3}-\omega_{2},\quad \Delta_{2}=\omega_{3}, 
\end{equation}
so the solution is obtained to be 
\begin{eqnarray}
\Phi(t)&=&\mbox{exp}(-it\tilde{H}_{IV})\Phi(0)=        \nonumber \\
&=&
\left(
  \begin{array}{ccc}
  \frac{g_{3}^{2}\mbox{cos}(\sqrt{g_{3}^{2}+g_{2}^{2}}t)+g_{2}^{2}}
  {g_{3}^{2}+g_{2}^{2}}& 
  \frac{g_{3}g_{2}\mbox{cos}(\sqrt{g_{3}^{2}+g_{2}^{2}}t)-g_{3}g_{2}}
  {g_{3}^{2}+g_{2}^{2}}&
 -i\frac{g_{3}\mbox{sin}(\sqrt{g_{3}^{2}+g_{2}^{2}}t)}
  {\sqrt{g_{3}^{2}+g_{2}^{2}}}                                         \\
  \frac{g_{3}g_{2}\mbox{cos}(\sqrt{g_{3}^{2}+g_{2}^{2}}t)-g_{3}g_{2}}
  {g_{3}^{2}+g_{2}^{2}}&
  \frac{g_{2}^{2}\mbox{cos}(\sqrt{g_{3}^{2}+g_{2}^{2}}t)+g_{3}^{2}}
  {g_{3}^{2}+g_{2}^{2}}&
  -i\frac{g_{2}\mbox{sin}(\sqrt{g_{3}^{2}+g_{2}^{2}}t)}
  {\sqrt{g_{3}^{2}+g_{2}^{2}}}                                         \\
 -i\frac{g_{3}\mbox{sin}(\sqrt{g_{3}^{2}+g_{2}^{2}}t)}
  {\sqrt{g_{3}^{2}+g_{2}^{2}}}&
 -i\frac{g_{2}\mbox{sin}(\sqrt{g_{3}^{2}+g_{2}^{2}}t)}
  {\sqrt{g_{3}^{2}+g_{2}^{2}}}&
  \mbox{cos}(\sqrt{g_{3}^{2}+g_{2}^{2}}t)
  \end{array}
\right)\Phi(0).
\end{eqnarray}

As a result the solution we are looking for is 
\begin{eqnarray}
\Psi(t)&=&\mbox{e}^{-itE_{0}}
\left(
  \begin{array}{ccc}
  1 & &                                                           \\
    & \mbox{e}^{-i(\phi_{3}+\omega_{3}t-\phi_{2}-\omega_{2}t)} &  \\
    & & \mbox{e}^{-i(\phi_{3}+\omega_{3}t)}
  \end{array}
\right)\times           \nonumber \\
&&
\left(
  \begin{array}{ccc}
  \frac{g_{3}^{2}\mbox{cos}(\sqrt{g_{3}^{2}+g_{2}^{2}}t)+g_{2}^{2}}
  {g_{3}^{2}+g_{2}^{2}}& 
  \frac{g_{3}g_{2}\mbox{cos}(\sqrt{g_{3}^{2}+g_{2}^{2}}t)-g_{3}g_{2}}
  {g_{3}^{2}+g_{2}^{2}}&
 -i\frac{g_{3}\mbox{sin}(\sqrt{g_{3}^{2}+g_{2}^{2}}t)}
  {\sqrt{g_{3}^{2}+g_{2}^{2}}}                                         \\
  \frac{g_{3}g_{2}\mbox{cos}(\sqrt{g_{3}^{2}+g_{2}^{2}}t)-g_{3}g_{2}}
  {g_{3}^{2}+g_{2}^{2}}&
  \frac{g_{2}^{2}\mbox{cos}(\sqrt{g_{3}^{2}+g_{2}^{2}}t)+g_{3}^{2}}
  {g_{3}^{2}+g_{2}^{2}}&
  -i\frac{g_{2}\mbox{sin}(\sqrt{g_{3}^{2}+g_{2}^{2}}t)}
  {\sqrt{g_{3}^{2}+g_{2}^{2}}}                                         \\
 -i\frac{g_{3}\mbox{sin}(\sqrt{g_{3}^{2}+g_{2}^{2}}t)}
  {\sqrt{g_{3}^{2}+g_{2}^{2}}}&
 -i\frac{g_{2}\mbox{sin}(\sqrt{g_{3}^{2}+g_{2}^{2}}t)}
  {\sqrt{g_{3}^{2}+g_{2}^{2}}}&
  \mbox{cos}(\sqrt{g_{3}^{2}+g_{2}^{2}}t)
  \end{array}
\right)\Phi(0).
\end{eqnarray}

For the latter use we set 
\begin{eqnarray}
\widetilde{U}_{6}(t,0)&=&\mbox{e}^{-itE_{0}}
\left(
  \begin{array}{ccc}
  1 & &                                                           \\
    & \mbox{e}^{-i(\phi_{3}+\omega_{3}t-\phi_{2}-\omega_{2}t)} &  \\
    & & \mbox{e}^{-i(\phi_{3}+\omega_{3}t)}
  \end{array}
\right)\times           \nonumber \\
&&
\left(
  \begin{array}{ccc}
  \frac{g_{3}^{2}\mbox{cos}(\sqrt{g_{3}^{2}+g_{2}^{2}}t)+g_{2}^{2}}
  {g_{3}^{2}+g_{2}^{2}}& 
  \frac{g_{3}g_{2}\mbox{cos}(\sqrt{g_{3}^{2}+g_{2}^{2}}t)-g_{3}g_{2}}
  {g_{3}^{2}+g_{2}^{2}}&
 -i\frac{g_{3}\mbox{sin}(\sqrt{g_{3}^{2}+g_{2}^{2}}t)}
  {\sqrt{g_{3}^{2}+g_{2}^{2}}}                                         \\
  \frac{g_{3}g_{2}\mbox{cos}(\sqrt{g_{3}^{2}+g_{2}^{2}}t)-g_{3}g_{2}}
  {g_{3}^{2}+g_{2}^{2}}&
  \frac{g_{2}^{2}\mbox{cos}(\sqrt{g_{3}^{2}+g_{2}^{2}}t)+g_{3}^{2}}
  {g_{3}^{2}+g_{2}^{2}}&
  -i\frac{g_{2}\mbox{sin}(\sqrt{g_{3}^{2}+g_{2}^{2}}t)}
  {\sqrt{g_{3}^{2}+g_{2}^{2}}}                                         \\
 -i\frac{g_{3}\mbox{sin}(\sqrt{g_{3}^{2}+g_{2}^{2}}t)}
  {\sqrt{g_{3}^{2}+g_{2}^{2}}}&
 -i\frac{g_{2}\mbox{sin}(\sqrt{g_{3}^{2}+g_{2}^{2}}t)}
  {\sqrt{g_{3}^{2}+g_{2}^{2}}}&
  \mbox{cos}(\sqrt{g_{3}^{2}+g_{2}^{2}}t)
  \end{array}
\right).
\end{eqnarray}

\vspace{5mm}

\begin{flushleft}
\begin{Large}
{\bf Unitary Transformation of type {VII}}
\end{Large}
\end{flushleft}

The Hamiltonian is 
\begin{eqnarray}
\label{eq:hamiltonian-7-g}
H_{VII}&=&
\left(
  \begin{array}{ccc}
   E_{0} & g_{1}\mbox{e}^{i(\phi_{1}+\omega_{1}t)} & 
   g_{3}\mbox{e}^{i(\phi_{3}+\omega_{3}t)}               \\
   g_{1}\mbox{e}^{-i(\phi_{1}+\omega_{1}t)} & E_{1} & 
   g_{2}\mbox{e}^{i(\phi_{2}+\omega_{2}t)}               \\
   g_{3}\mbox{e}^{-i(\phi_{3}+\omega_{3}t)} & 
   g_{2}\mbox{e}^{-i(\phi_{2}+\omega_{2}t)} & E_{2}
  \end{array}
\right)              \nonumber \\
&=&E_{0}{\bf 1}_{3}+
\left(
  \begin{array}{ccc}
   0 & g_{1}\mbox{e}^{i(\phi_{1}+\omega_{1}t)} & 
   g_{3}\mbox{e}^{i(\phi_{3}+\omega_{3}t)}                  \\
   g_{1}\mbox{e}^{-i(\phi_{1}+\omega_{1}t)} & \Delta_{1} & 
   g_{2}\mbox{e}^{i(\phi_{2}+\omega_{2}t)}                  \\
   g_{3}\mbox{e}^{-i(\phi_{3}+\omega_{3}t)} & 
   g_{2}\mbox{e}^{-i(\phi_{2}+\omega_{2}t)} & \delta_{2}
  \end{array}
\right).
\end{eqnarray}
Here we assume the following 
\[
\omega_{3}=\omega_{1}+\omega_{2},
\]
which is a kind of consistency condition. 

We would like to solve the Schr{\" o}dinger equation 
\begin{equation}
\label{eq:schrodinger-equation-7-g}
i\frac{d}{dt}\Psi=H_{VII}\Psi.
\end{equation}

If we note the following decomposition 
\begin{eqnarray}
&&\left(
  \begin{array}{ccc}
   0 & g_{1}\mbox{e}^{i(\phi_{1}+\omega_{1}t)} & 
   g_{3}\mbox{e}^{i(\phi_{3}+\omega_{3}t)}                  \\
   g_{1}\mbox{e}^{-i(\phi_{1}+\omega_{1}t)} & \Delta_{1} & 
   g_{2}\mbox{e}^{i(\phi_{2}+\omega_{2}t)}                  \\
   g_{3}\mbox{e}^{-i(\phi_{3}+\omega_{3}t)} & 
   g_{2}\mbox{e}^{-i(\phi_{2}+\omega_{2}t)} & \Delta_{2}
  \end{array}
\right)
=               \nonumber \\
&&\left(
  \begin{array}{ccc} 
  1 & &                                                           \\
    & \mbox{e}^{-i(\phi_{1}+\omega_{1}t)} &                       \\
    & & \mbox{e}^{-i(\phi_{1}+\omega_{1}t+\phi_{2}+\omega_{2}t)}
  \end{array}
\right)
\left(
  \begin{array}{ccc}
   0 & g_{1} & g_{3}\mbox{e}^{i(\phi_{3}-\phi_{1}-\phi_{2})}           \\
   g_{1} & \Delta_{1} & g_{2}                                          \\
   g_{3}\mbox{e}^{-i(\phi_{3}-\phi_{1}-\phi_{2})} & g_{2} & \Delta_{2}
  \end{array}
\right) \times     \nonumber \\
&&
\left(
  \begin{array}{ccc}
  1 & &                                                          \\
    & \mbox{e}^{i(\phi_{1}+\omega_{1}t)} &                       \\
    & & \mbox{e}^{i(\phi_{1}+\omega_{1}t+\phi_{2}+\omega_{2}t)} 
  \end{array}
\right) 
\end{eqnarray}
and define 
\begin{equation}
\Phi=\mbox{e}^{itE_{0}}
\left(
  \begin{array}{ccc}
  1 & &                                                           \\
    & \mbox{e}^{i(\phi_{1}+\omega_{1}t)}  &                       \\
    & & \mbox{e}^{i(\phi_{1}+\omega_{1}t+\phi_{2}+\omega_{2}t)}
  \end{array}
\right)\Psi 
\end{equation}
, then it is not difficult to see 
\begin{equation}
i\frac{d}{dt}\Phi=
\left(
  \begin{array}{ccc}
   0 & g_{1} & g_{3}\mbox{e}^{i(\phi_{3}-\phi_{1}-\phi_{2})}      \\
   g_{1} & \Delta_{1}-\omega_{1} & g_{2}                          \\
   g_{3}\mbox{e}^{-i(\phi_{3}-\phi_{1}-\phi_{2})} & g_{2} & 
   \Delta_{2}-\omega_{1}-\omega_{2}
  \end{array}
\right)\Phi
\equiv 
\tilde{H}_{VII}\Phi.
\end{equation}

Here we set the resonance condition 
\begin{equation}
\label{eq:resonance-7-g}
 \Delta_{1}=\omega_{1}\quad \mbox{and}\quad 
 \Delta_{2}=\omega_{1}+\omega_{2}=\omega_{3},
\end{equation}
and moreover the phase condition 
\begin{equation}
\label{eq:phase-7-g}
 \phi_{3}=\phi_{1}+\phi_{2}.
\end{equation}
To obtain the solution we must calculate 
\begin{eqnarray}
\mbox{exp}(-it\tilde{H}_{VII})=
\mbox{exp}
\left\{-it
\left(
  \begin{array}{ccc}
   0 & g_{1} & g_{3}   \\
   g_{1} & 0 & g_{2}   \\
   g_{3} & g_{2} & 0
  \end{array}
\right)
\right\}.
\end{eqnarray}
However we cannot do it (this calculation seems to be very difficult), so 
we present it as 

\begin{flushleft}
{\bf Problem} \quad Calculate it. 
\end{flushleft}

\newpage 

\noindent{\em Acknowledgment.}\\
K. Fujii wishes to thank Akira Asada and Kunio Funahashi for some useful 
suggestions.

\vspace{10mm}


\end{document}